\documentclass[useAMS,usenatbib]{mnras}
\usepackage{epsfig}
\usepackage{latexsym} 
\usepackage{amsmath}
\usepackage{graphicx}

\title[Correlations between DIBs]{The well correlated DIBs at $\lambda\lambda$
6196, 6614 \AA\ and their possible companions.}

\author[A. Bondar]{A. Bondar$^{1}$\thanks{E-mail: arctur.ab@gmail.com} \\ 
$^{1}$ International Center for Astronomical and Medico-Ecological Research,
Zabolotnoho Str. 27, Kyiv, Ukraine \\}

\date{Last updated 2020 May 29; in original form 2019 Dec 26}

\pubyear{2020}

\begin{document}
\label{firstpage}
\pagerange{\pageref{firstpage}--\pageref{lastpage}}
\maketitle
\newcommand{\ew}{equivalent width}
\newcommand{\ce}{E$_{\rm B-V}$}
\def\sp{\hspace{0.33pt}`$\bullet$'}
\def\scz{\scriptsize}
\def\icm{$cm^{-1}$}
\def\deltnu{$\Delta\tilde{\nu}$}
\def\tnu{$\tilde{\nu}$}
\def\lrar{$\leftrightarrow$}

\begin{abstract}

Considered here is the interrelation between five diffuse
interstellar bands (DIBs): $\lambda\lambda$ 5545, 6113, 6196, 6445
and 6614 \AA. Two DIBs ($\lambda$~6196 and $\lambda$~6614~\AA) have
already been known as well correlated with each other; their relation
with three other, weaker bands, was investigated for the first
time. To accomplish this task high-resolution spectra 
($\lambda/\delta\lambda\approx$100,000) with high signal-to-noise
ratio (S/N) of 54 hot O~-- B stars with reddening 0.12~-- 1.45
mag were used. Analysis of measured equivalent widths has allowed to
establish linear dependencies and evaluate linear correlation
coefficients as high as 0.968~-- 0.988 between the intensities of these
five DIBs. Such a degree of correlation may indicate their common
origin. Several spacings in wavenumbers found between these DIBs
correspond to the energies of vibrational transitions in some PAHs
resulting in IR emissions at $\lambda\lambda$~16.4, 11.3, 7.7, 6.2 and
3.3 \micron.

\end{abstract}


\section{Introduction}

During nearly a century astrophysicists have been searching for a key
to understanding of the origin of diffuse interstellar bands. A great
deal of works dedicated to the DIBs problem directly concerns the
subject of interrelation between DIBs as well as of the nature of
their carriers. So, \cite{Hb75} considered 39 diffuse bands in
spectra of 56 hot stars of diverse spectral types (from O6 till A5).
17 DIBs showed quite a good correlation between their strengths~---
even better than either with color excess. He did not find any
regularities in the wavenumbers of the analysed DIBs that might be
considered, e.\,g. as a series of vibrational transitions. As DIBs
carrier the author suggested a constituent of very small interstellar
grains $\approx 300$~\AA\ in radius. 

\cite{Ch86} investigated several DIBs in spectra of stars from three
associations. A special attention was given to the shapes of profiles
and correlations between, so-called, ``major'' bands $\lambda\lambda$
5780, 5797, 6195, 6203, 6270 and 6284 \AA. Basing on obtained results
the authors concluded that DIBs could rather arise due to vibrational
transitions in large molecules than on interstellar dust grains. To
the same inferences Chlewicki et al. came after the study of
correlations between narrow and broad DIBs \citep[see][]{Ch87}. 
There authors put forward several arguments against the solid state
origin hypothesis of DIBs, such as constancy of their central
wavelengths, differences in shapes of profiles amid strongly
correlated bands, existence of narrow diffuse bands and correlations
between narrow and broad DIBs, that suggests close relation of their
carriers. But as the hypothesis excludes appearance of the narrow DIBs
so solid particles should not also be responsible for existence of the
broad ones.

\cite{KW87} analysing intensities and central depths of some diffuse
bands in spectra of reddened early B stars (\ce\ = 0.16~-- 0.42)
detected three probable families of DIBs, most evident in the spectrum
of $\zeta$~Per. The authors supposed existence of three different
carriers for those DIBs. 

Concept of DIB families got its continuation in the paper by
\cite{KS95} and in quite recent works, for instance, by \cite{Fr11,
Vs11, Kr19}, where spectra of selected objects were classified
accordingly to intensity ratios of two strong DIBs $\lambda\lambda$
5797 and 5780 \AA\ as `$\sigma$' and `$\zeta$' interstellar cloud
types~--- called after sightlines towards $\sigma$~Sco and
$\zeta$~Oph. 

But \cite{Ca97} besides `$\sigma$' and `$\zeta$' cloud types
considered CS (circumstellar) and the Orion type clouds.  Their
classification of DIB families is based on differences in ionization
potentials (an ionization hypothesis) of DIB carriers~--- as probably
large carbon-bearing molecules. In their opinion DIBs that correlate
well with each other have carriers with similar ionization potentials
and vise verse: poorly correlated DIBs should have carriers with
significant differences in ionization and recombination properties. On
the whole the authors support molecular hypothesis of DIB carriers as
large carbonaceous molecules and explain existence of `$\zeta$',
`$\sigma$' and Orion clouds as a sequence in intensity of local
UV-field: from low in `$\zeta$' through strong in `$\sigma$' to
extreme UV radiation in Orion cloud types.

\cite{Hb88} described six weak features nearly uniformly spaced
at 35 cm$^{-1}$ between 6779~-- 6860 \AA. He speculated about
rotational or vibrational progressions with spacings at 35 cm$^{-1}$
level, but did not find any plausible carrier for this group of bands.

\cite{HbL91} explored the systematics in spacings of more than hundred of
DIBs. Assuming that in interstellar medium only the lowest energy
levels (no excitation) of possible complex molecules are involved into
DIBs production, they sought for vibrational sequences of the type $0
\rightarrow {\it v}'$. As a result, from 50 sequences of four or more
DIBs, found by their method, eight were considered as reliable: thus
some of the bands in these sequences could be related to one common
carrier. The authors had taken into account quite a big number of
known bands and concluded that DIBs should be produced by a mixture of
different carriers.  

\cite{Mo99} tested 10 DIBs in spectra of 62 stars and found good
correlations between several of them: 6614 $\leftrightarrow$ 5780,
6614 $\leftrightarrow$ 5797, 6196 $\leftrightarrow$ 5780, 6196
$\leftrightarrow$ 5797, 6196 $\leftrightarrow$ 6379, 5797
$\leftrightarrow$ 5850, 5797 $\leftrightarrow$ 6379 with correlation
coefficients ranging 0.90 -- 0.93. But there was one strict
correlation between equivalent widths of the DIBs $\lambda\lambda$~
6614 and 6196~\AA, with correlation coefficient R = 0.98 --- the only
case for which a common carrier was quite real.

Later detailed study of this pair based on a significant sample of
spectra (114 targets) was carried out by \cite{MC10}. The authors got
excellent correlation between these two DIBs (R = 0.986), but final
conclusion about their common carrier was left for future
investigations due to difference in forms of their profiles.

Afterwards some works pertained to matter of correlation between
different DIBs, mainly the strong ones~--- $\lambda\lambda$~5780,
5797, 6196, 6204, 6283 and 6614~\AA. Thus, \cite{Fr11} found moderate
(R = 0.87~-- 0.98) correlations between these bands besides two, much
better correlated (R~= 0.99) bands $\lambda\lambda$ 6196.0,
6613.6~\AA. In \cite{Vs11} the most correlated pair was DIBs at 6379
and 6614~\AA\ (R~= 0.92); the DIBs at 6196 and 6614~\AA\ demonstrated
quite a low degree of correlation, with R = 0.80. Similar result (R =
0.89) for the two last was obtained by \cite{Ba16}. These researchers
also found a break in the monotonic relation between these diffuse
bands at low values of intensities: EW(6614) $\approx$ 20.0~--
40.0~m\AA, EW(6196) $<$ 10~m\AA. Though \cite{Kr16} using spectra of
much higher resolution confirmed tight correlation between these two
diffuse bands, as it was presented in \cite{MC10}.

In this paper the author presents a new study of interrelation between
the well-correlated DIBs $\lambda\lambda$ 6195.95 and 6613.58 \AA\ and
three relatively weak bands: $\lambda\lambda$ 5545.00, 6113.18 and
6445.22 \AA. As the differences in wavenumbers for these five DIBs are
close to vibrational energies in molecules an attempt to find suitable
molecules, candidates in DIB carriers was done. The wavelengths for
these diffuse bands were taken from work by \cite{Bo12} and further in
the text accepted as $\lambda$~5545, 6113, 6196, 6445 and 6614 DIBs.

\section{Objects and spectral data.}
\subsection{The objects.}
 
The selected objects (54 in total) cover wide range of spectral
classes from O5 to B9.5 with several Wolf-Rayet stars. Many of them
are members of well-known stellar associations, such as Per\,OB2,
Sgr\,OB1, Sco\,OB2, $\rho$~Oph cloud and others, so they present
regions of interstellar medium with variant physical conditions.
These stars have reddening from moderate to strong: \ce = 0.12~--
1.45 mag. The main characteristics of the selected targets are
presented in the Table~\ref{targets}. Intrinsic values of (B$-$V)$_0$
for calculating \ce\ in the 5th column of the Table~\ref{targets} were
taken from work by \cite{Pa93}.

One of the selection criteria for the objects in the
Table~\ref{targets} is the absence of a splitting in interstellar atomic
or molecular lines visible in their spectra, i.\,e. a single cloud on
the sightline towards a star. Or at least, if such a splitting exists,
a secondary component should have much lesser intensity in comparison
with a main component, the strongest one. This criterion was not
strictly kept as it is quite a difficult task in case of high
resolution spectra. The 1st column of the Table~\ref{targets} clearly
demonstrates that 25\% of sightlines (marked with \sp\ sign) have
more than one interstellar cloud. All the other objects in the
Table~\ref{targets} either do not have any signs of splitting in the
potassium \ion{K}{I} line or can have slight asymmetries in this line.
A special case is the star HD\,169454 as its spectrum reveals two
evident interstellar clouds towards it.

As a rule verification was done on the interstellar line of potassium
\ion{K}{I} at $\lambda$~7699~\AA, or in case of its absence
(HARPS spectra) on other molecular (mainly CH at $\lambda$~4300~\AA)
or atomic ({\rm \ion{Ca}{I}} at $\lambda$~4227~\AA) interstellar
lines.

\subsection{The spectra.}

The used sample of spectral data was collected in different years with
three spectrographs: mostly HARPS (2007~yr., program ID 078.C-0403(A))
and UVES (2009 and 2015 yrs., program IDs 082.C-0566(A),
092.C-0019(A)), see respective acronyms (`H', `U') in the last column
of the Table~\ref{targets}. Also, spectra obtained with HARPS and UVES
spectrographs from ESO science archive have been used. They include
the raw spectra as well as the reduced ``Phase3'' ones. Spectra of
five objects were obtained with BOES spectrograph \citep{Km07}, letter
`B' in the last column of the Table~\ref{targets}. All the considered
spectra are of a high resolution, $\lambda/\delta\lambda \approx$
90,000~-- 120,000.  Measured S/N in vicinity of the selected DIBs
ranges from $\approx$260 to 2100. Minimum and maximum S/N (with noted
spectrum of the star) and S/N mean values for all the spectra in the
data set are next: 

{\small
\begin{itemize}
\item DIB~5545: {\bf 430} (HD\,147889)~-- {\bf 2010} (HD\,159176), $<\bf 790>$;
\item DIB~6113: {\bf 270} (HD\,144470)~-- {\bf 2080} (HD\,142468), $<\bf 685>$;
\item DIB~6196: {\bf 330}\hspace{16pt} (15~Sgr)~-- {\bf 1620}\hspace{20pt} ($\sigma$~Sco), $<\bf 725>$;
\item DIB~6445: {\bf 260} (HD\,185859)~-- {\bf 1920} (HD\,142468), $<\bf 615>$;
\item DIB~6614: {\bf 305} (HD\,203532)~-- {\bf 1750} \hspace{5pt}(HD\,63804), $<\bf 650>$.
\end{itemize}
}

\begin{table*}
\centering
\caption{Main characteristics of the selected objects: visual
magnitude (V), spectral type/luminosity (Sp/L), reddening (\ce),
spectrograph used (Inst.).}
\label{targets}
\begin{tabular}{rclllr} 
	Name$^a$& Type$^b$& V, {\scz mag}& Sp/L &\ce, {\scz mag}& Inst.$^c$ \\
\hline
	HD 14956  & s$*$b & 7.20 & B1.5Ia    & 0.89 & B \\
	HD 23016  & Be$*$ & 5.69 & B9Vne     & 0.12 & U \\
	  omi Per &  $**$ & 3.83 & B1III     & 0.27 & U, B \\
	  zet Per &  V$*$ & 2.85 & B1Ib      & 0.29 & U \\
     62 Tau\sp$^d$&   $*$ & 6.34 & B3V       & 0.37 & U \\
	HD 34078  & Or$*$ & 5.96 & O9.5V     & 0.49 & U, B \\
	HD 37367\sp&   $*$ & 5.98 & B2IV-V    & 0.34 & U \\
	LS 719    & s$*$b & 9.19 & B3Ia      & 0.99 & U \\
	HD 63804  & s$*$b & 7.78 & B9.5Ia+(e)& 1.05 & U \\
	HD 73882  & EB$*$ & 7.19 & O8.5IV    & 0.68 & U \\
	HD 76341  & Em$*$ & 7.16 & O9.2IV    & 0.50 & U \\
	HD 78344  &   $*$ & 9.00 & O9.5/B0(Ib)&1.34 & U \\
	HD 80077  & s$*$b & 9.00 & B2Ia+e    & 1.45 & U \\
	HD 91824\sp& SB$*$ & 8.14 & O7V((f))z & 0.26 & U \\
	Tr~16~112\sp& SB$*$ & 9.22 & O6V((fc)) & 0.61 & U \\
	HD 97253\sp& SB$*$ & 7.09 & O5III(f)  & 0.37 & U \\
	HD 100099 &   $*$ & 8.08 & O8/9      & 0.37 & U \\
	HD 110432 & Be$*$ & 5.31 & B0.5IVpe  & 0.47 & U \\
	HD 142468 &  V$*$ & 7.90 & B1Ia/Iab  & 0.77 & U \\
	HD 144217 & SB$*$ & 2.62 & B1V       & 0.16 & U, H \\
	HD 144470 & bC$*$ & 3.97 & B1V       & 0.19 & U \\
	HD 145502 &  $**$ & 4.00 & B2V       & 0.25 & U \\
	  sig~Sco & bC$*$ & 2.89 & O9.5(V)+B7(V)& 0.35 & H \\
	HD 147888 &  $**$ & 6.74 & B3/4V     & 0.48 & U \\
	HD 147889 & pr$*$ & 7.90 & B2III/IV  & 1.09 & U, H \\
	HD 147933 &   $*$ & 5.05 & B2/3V     & 0.43 & H \\
	  chi~Oph & Be$*$ & 4.43 & B2Vne     & 0.49 & U, H \\
	HD 148688\sp& s$*$b & 5.39 & B1Iaeqp   & 0.52 & U \\
	  zet~Oph & Be$*$ & 2.56 & O9.2IVnn  & 0.29 & U \\
	HD 151932\sp& WR$*$ & 6.51 & WN7h      & 0.57 & U \\
	HD 152003 & s$*$b & 7.08 & O9.7IabNwk& 0.61 & U \\
	HD 152233\sp& SB$*$ & 6.59 & O6II(f)   & 0.39 & H \\
	HD 152235\sp& s$*$b & 6.38 & B0.5Ia    & 0.74 & U \\
	HD 152249 & s$*$b & 6.45 & OC9Iab    & 0.42 & U \\
	HD 152270 & WR$*$ & 6.59 & WC7+O5-8  & 0.50 & U \\
	HD 154090\sp& s$*$b & 4.87 & B2Iab     & 0.45 & H \\
	HD 154445 &   $*$ & 5.61 & B1V       & 0.37 & U \\
	HD 155450 &   $*$ & 6.00 & B1Ib      & 0.21 & H \\
	HD 159176 & SB$*$ & 5.70 & O7V((f))+O7V((f))& 0.34 & U \\
	HD 161056\sp&   $*$ & 6.32 & B3II/III  & 0.59 & U \\
	V2052~Oph & bC$*$ & 5.82 & B3III     & 0.21 & U \\
	HD 163800 &   $*$ & 7.00 & O7.5III((f))& 0.57 & U, H \\
	   9~Sgr  & Em$*$ & 5.97 & O4V((f))z & 0.35 & U \\
	  15~Sgr  & s$*$b & 5.37 & O9.7Iab   & 0.29 & U \\
	HD 167771 & SB$*$ & 6.54 & O7III((f))+O8III & 0.37 & U \\
	HD 169454\sp& s$*$b & 6.71 & B1Ia      & 1.10 & U \\
	HD 170740 & SB$*$ & 5.72 & B2/3II    & 0.40 & U \\
	  20~Aql  &  V$*$ & 5.34 & B2/3II    & 0.31 & U, H \\
	HD 184915 &   $*$ & 4.96 & B0.5IIIn  & 0.22 & U \\
	HD 185418\sp&   $*$ & 7.49 & B0.5V     & 0.39 & U \\
	HD 185859 & Em$*$ & 6.52 & B0.5Iae   & 0.59 & U \\
	HD 193793\sp& WR$*$ & 6.85 & WC7p+O5   & 0.69 & B \\
	HD 203532 &   $*$ & 6.38 & B3IV/V    & 0.31 & U \\
	HD 228712 & s$*$b & 8.67 & B0.5Ia    & 1.34 & B \\
\hline  
\multicolumn{6}{l}{Notes.} \\
\multicolumn{6}{l}{$^a$ HD number, Bayer or other designation.} \\
\multicolumn{6}{p{275pt}}{$^b$ Short type of an object (acronyms
adopted from SIMBAD): $*$~--- star, s$*$b~--- blue supergiant star,
$**$~--- double star, SB$*$~--- spectral binary, V$*$~--- variable
star, Or$*$~--- Variable Star of Orion Type, bC$*$~--- Variable Star
of beta Cep type, Be$*$~--- Be-type star, WR$*$~--- Wolf-Rayet star,
pr$*$~--- pre-main sequence star, Em$*$~--- emission-line star,
EB$*$~--- eclipsing binary.} \\
\multicolumn{6}{l}{$^c$ `B'~--- BOES, `H'~--- HARPS, `U'~--- UVES.} \\
\multicolumn{6}{p{275pt}}{$^d$ '\sp'\ sightlines with strong asymmetry or
signs of splitting in interstellar atomic \ion{K}{I} line at
$\lambda$~7699 \AA.} \\  
\end{tabular}
\end{table*}

So, the spectral intervals utilized in this work have high mean S/N
close to 700. 

\subsection{Choice of diffuse bands.}

DIBs for the current study were selected on suggestion that they
preserve their intensity ratios in the same manner as DIBs 6196 and
6614 \AA\ while changing sightlines in the Galaxy: {\small
EW(DIB)$_1$/EW(DIB)$_2\approx\,$EW(6196)$_1$/EW(6196)$_2$$\approx$
EW(6614)$_1$/EW(6614)$_2$}, where indices 1, 2 stand for two different
sightlines. The intensity ratios were calculated by division
intensities of DIBs in spectrum of an arbitrary star by those in the
spectrum of HD\,163800. The spectrum of this object was chosen as
reference for quite strong DIBs in it and because it is a simple star:
without signs of binarity, variability and so on, see the
Table~\ref{targets}.

At first EWs for ten stars were taken from work by \cite{Bo12}. Later
on the set of objects was widened to 54 stars. Initially number of the
selected DIBs reached 35, but after a thorough verification most of
them were rejected as implausible and also due to big uncertainties in
their EWs, their weakness or too noisy spectra. Thus only three bands
which demonstrated variations in \ew s in agreement with those in DIBs
$\lambda\lambda$~6196 and 6614 \AA\ in different directions in the
Galaxy had been left~--- $\lambda\lambda$ 5545, 6113 and 6445 \AA. 

\section{Data analysis and results.}
\subsection{Data reduction.}

Raw spectra were processed by the author either with IRAF \citep{IRAF}
or ESO UVES pipeline:
\url{http://www.eso.org/sci/software/pipelines/}, ``esorex'' and
``gasgano'' programs. A set of reduced spectra from HARPS (2007) and
UVES (2015) spectrographs was kindly given by Prof.~J.~Kre{\l}owski.

\begin{figure*}
\centering
  \begin{tabular}{@{}cc@{}}
    \includegraphics[width=.425\textwidth]{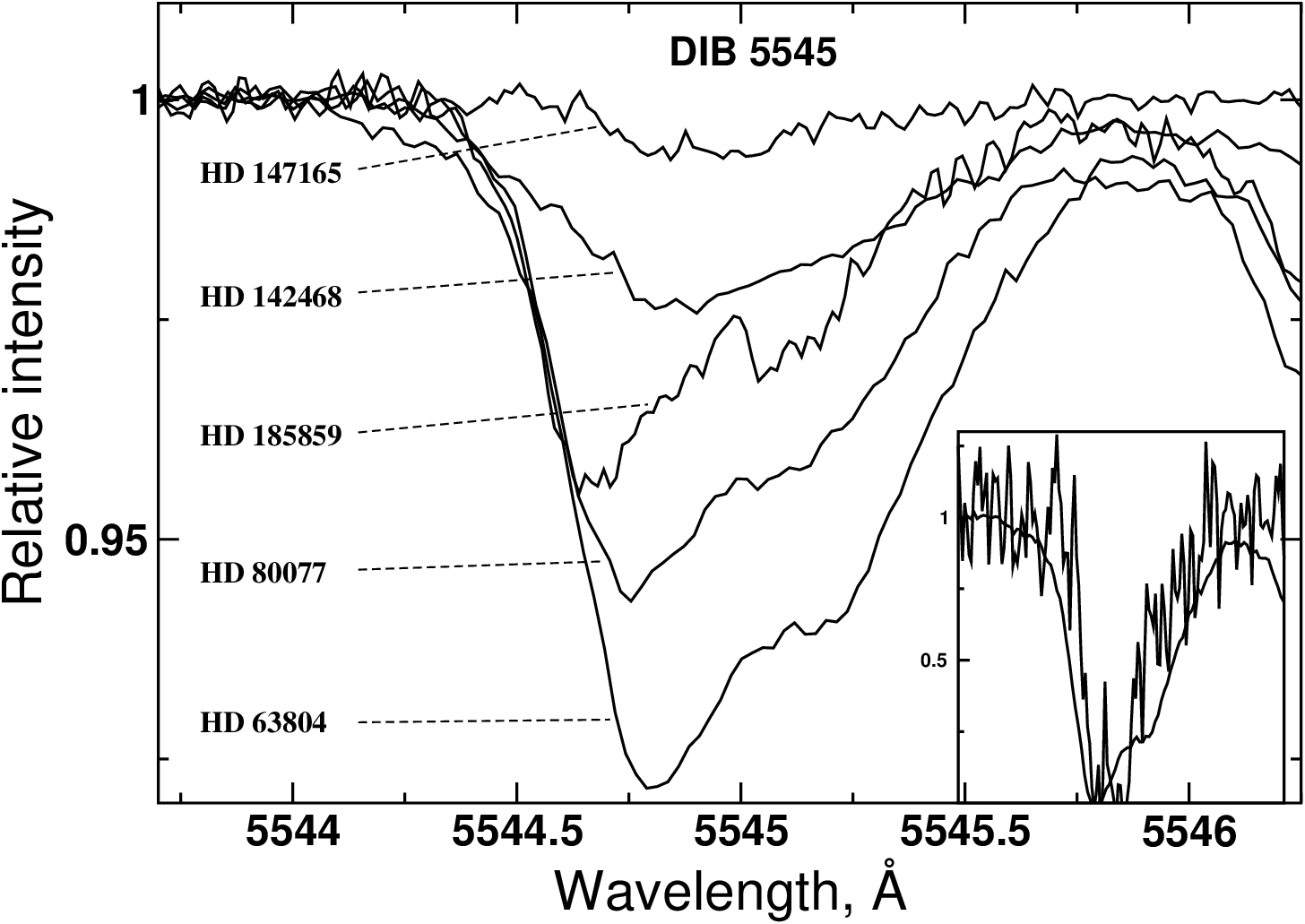} &
    \includegraphics[width=.425\textwidth]{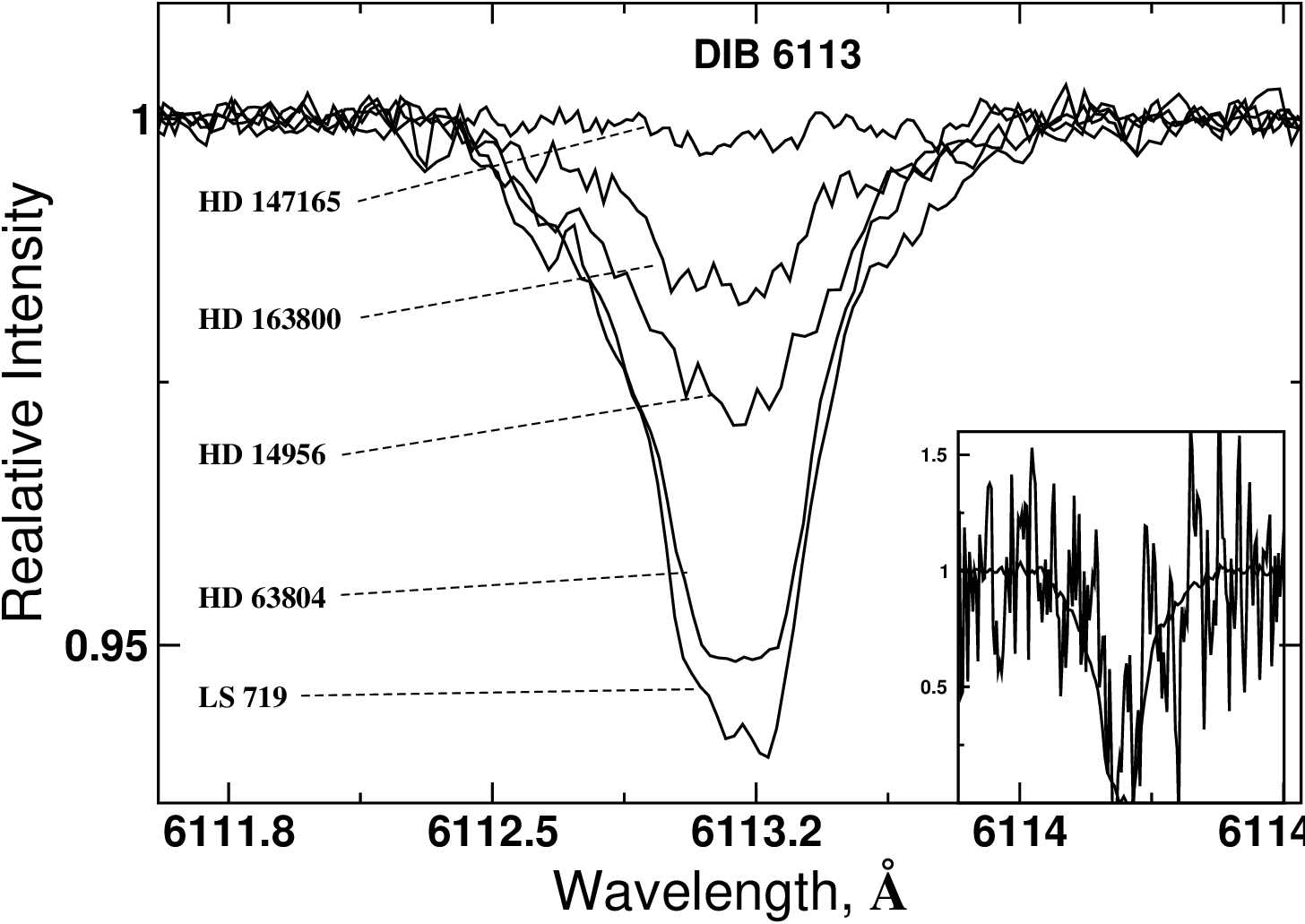} \\
    \includegraphics[width=.425\textwidth]{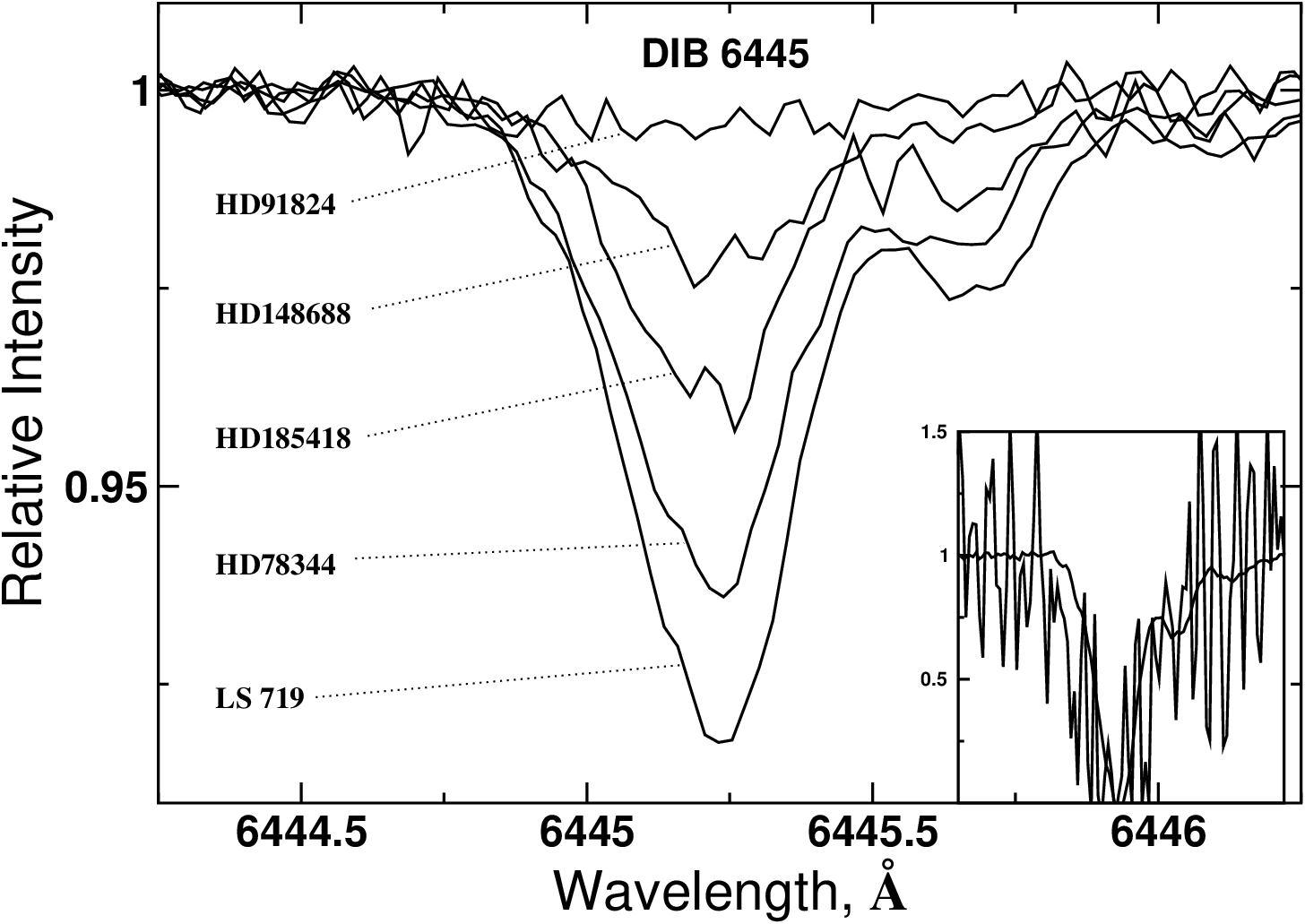} &
    \includegraphics[width=.425\textwidth]{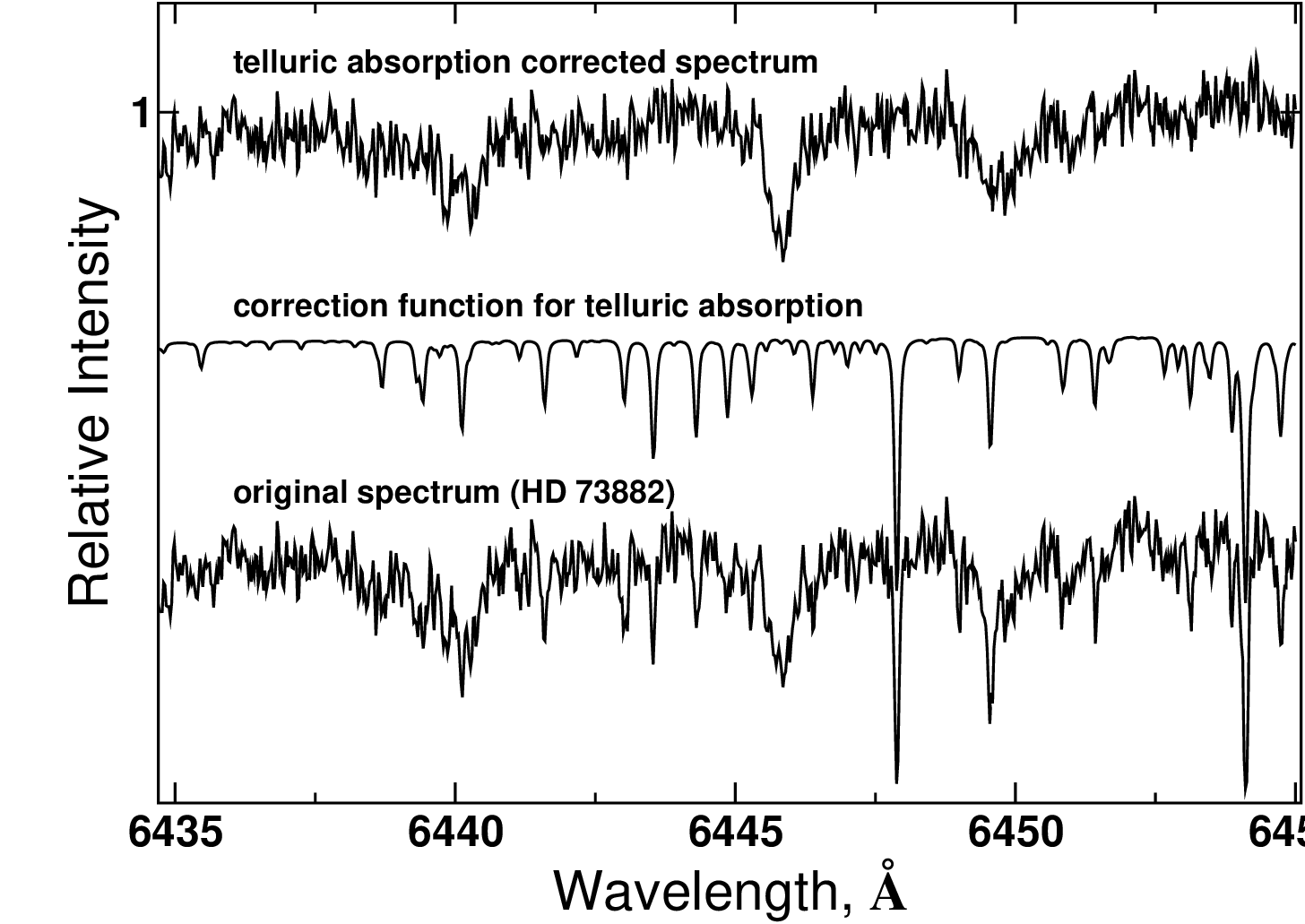} \\
    \includegraphics[width=.425\textwidth]{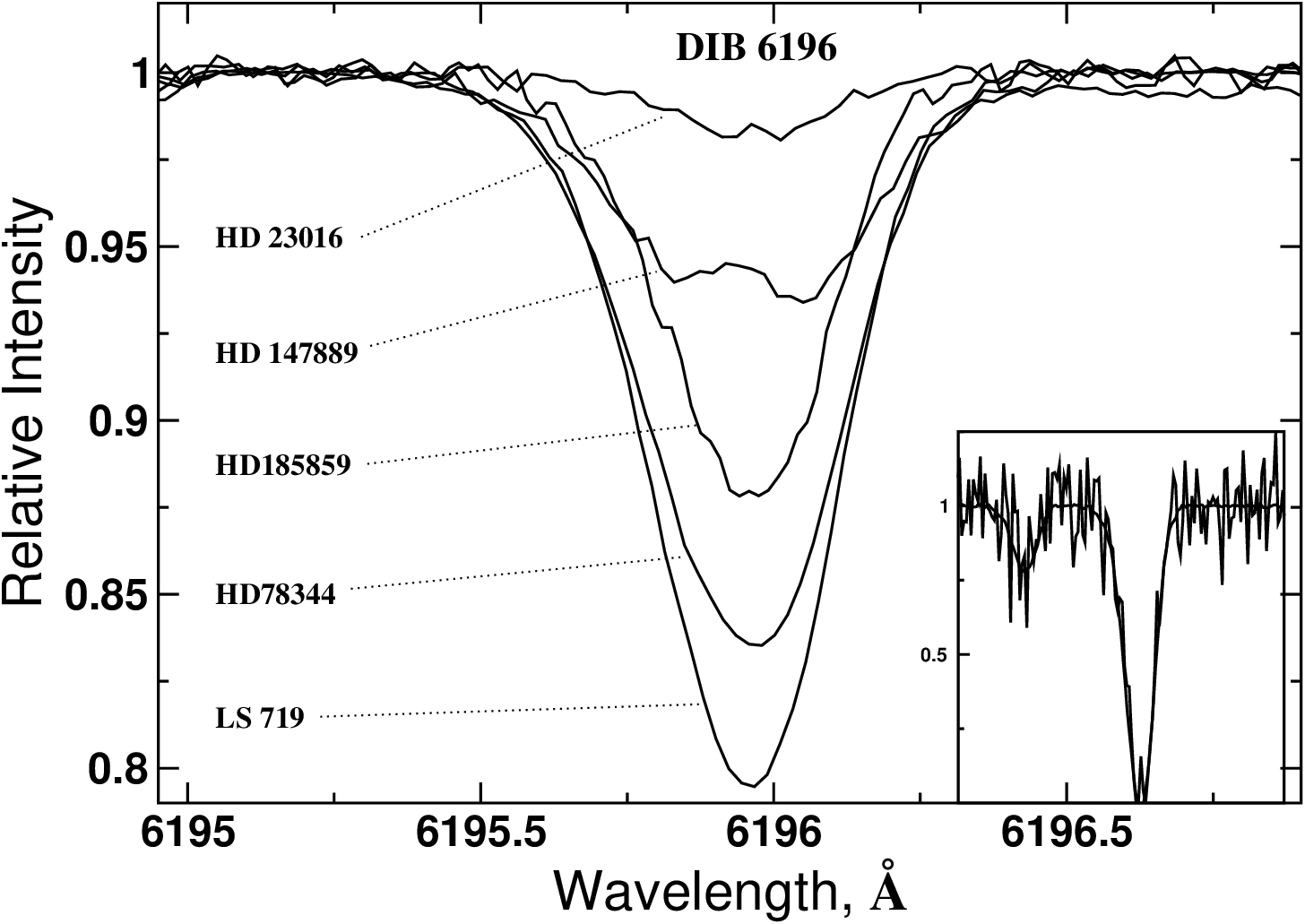} &
    \includegraphics[width=.425\textwidth]{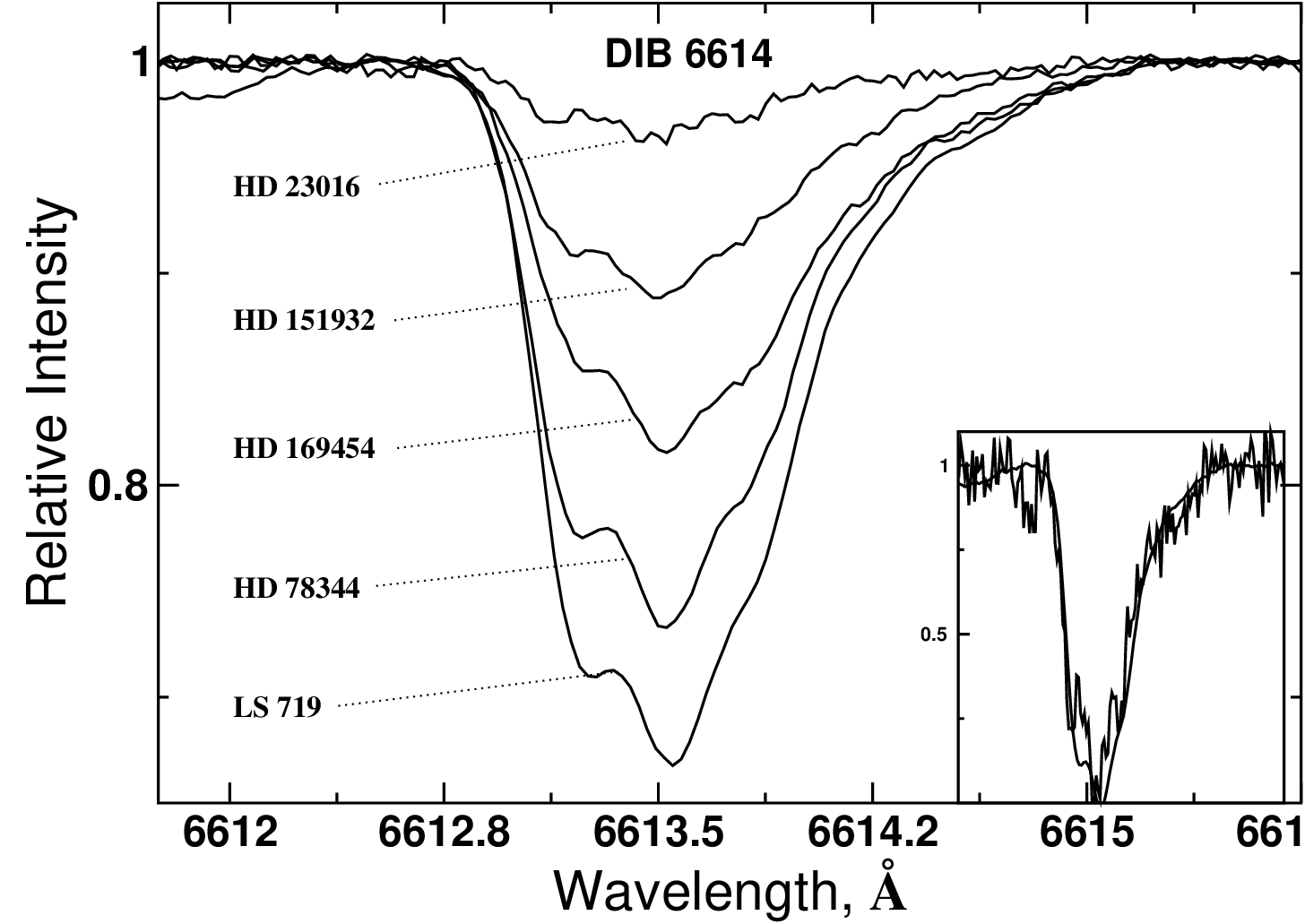} \\
  \end{tabular}
\caption{The selected DIBs of different intensities with
characteristic profiles. The insets below right depict comparison
between the weakest and strongest normalized profiles of the DIB. The
mid picture at the right panel demonstrates telluric lines removing
procedure for DIB at 6445 \AA.}
\label{profil}
\end{figure*}

\subsection{Analysis of the one-dimensional spectra.}

All the one-dimensional spectra were analysed with IRAF spectroscopic
package tasks. A special procedure for measurement of equivalent
widths with evaluation of uncertainties by method of \cite{Vo06} was
created on the basis of the IRAF ``splot'' task. Equivalent widths
have been measured by direct integration of DIB profiles.

Typical profiles of the $\lambda\lambda$~5545, 6113, 6196, 6445 and
6614 \AA\ DIBs ordered according to their intensities are depicted at
the Fig.~\ref{profil}. The insets below right, at every picture, show
comparison between the weakest and the strongest normalized profiles
of the DIB in the spectral data set. It can be seen that the profiles
mostly persist their forms well. Though some noticeable variations in
them, probably caused by specific local conditions in ISM or by blends
with stellar lines are visible: in width, see for example, the inset
for DIB 5545~\AA\ in the spectra of HD\,63804 and $\sigma$~Sco (the
strongest and the weakest cases) or appeared structure in the profile
of DIB~6196 \AA\ in the spectrum of HD\,147889. Profiles of the weak
DIBs in spectra with lower S/N can also be significantly distorted.

As to similarity of profiles for DIBs with common origin they do not
match well. Though profiles of DIBs $\lambda\lambda$ 6113 and 6196
\AA\ are quite alike. Common for three other bands is a redward
extension visible in their profiles (Fig.~\ref{profil}). 

Possible blends of the DIBs with stellar lines were verified on
synthetic spectra calculated with the programs ATLAS9 for Kurucz
model atmospheres \citep{Ku70, Ku92} and SYNTHE \citep{Sb04}  
according to the spectral type and luminosity of the selected stars.

Additionally, for DIB~6445 \AA\ a telluric lines cancellation
procedure was applied. For this goal spectrum of a hot star
($\alpha$~Vir) was used. But in many cases it has been realized with
``Molecfit'' package developed by \cite{molfit}. It was quite
effective, see e.\,g. mid picture at the right panel of the
Fig.~\ref{profil}.    

\begin{table*}
\centering
\caption{Equivalent widths with their uncertainties (m\AA) of the two
tightly correlated DIBs at $\lambda\lambda$ 6196, 6614 \AA\ and
diffuse bands likely related with them.}
\label{dib_ew}
\begin{tabular}{lrrrrr}
Star          &      5545    &    6113      &    6196      &    6445      &       6614     \\	
\hline
      hd14956 & 26.0$\pm$2.8 & 20.1$\pm$2.5 & 44.2$\pm$2.0 & 23.5$\pm$3.1 & 210.4$\pm$4.0 \\
      hd23016 &  4.0$\pm$1.8 &  2.9$\pm$3.0 &  6.5$\pm$2.1 &  3.1$\pm$3.5 &  34.8$\pm$4.9 \\
      omi~Per &  8.8$\pm$2.7 &  5.5$\pm$2.0 & 12.8$\pm$3.1 &  5.1$\pm$1.7 &  50.0$\pm$5.6 \\
      zet~Per &  8.7$\pm$2.2 &  7.3$\pm$4.9 & 13.4$\pm$2.1 &  6.1$\pm$4.0 &  60.8$\pm$9.3 \\
       62~Tau &  5.7$\pm$2.0 &  3.1$\pm$3.2 &  9.5$\pm$3.1 &  4.8$\pm$3.8 &  43.1$\pm$7.9 \\
      hd34078 & 12.9$\pm$3.4 &  4.7$\pm$3.7 & 18.9$\pm$2.7 &  5.9$\pm$3.1 &  57.9$\pm$5.6 \\
      hd37367 & 13.9$\pm$2.0 & 10.4$\pm$6.4 & 38.1$\pm$2.9 & 13.6$\pm$4.7 & 140.4$\pm$10.1 \\
        LS719 & 53.0$\pm$1.2 & 38.5$\pm$1.8 & 78.8$\pm$1.1 & 40.9$\pm$1.8 & 328.7$\pm$2.7 \\
      hd63804 & 63.3$\pm$1.3 & 33.5$\pm$2.4 & 81.2$\pm$1.1 & 36.6$\pm$1.9 & 316.2$\pm$1.9 \\
      hd73882 &  5.6$\pm$1.9 &  3.4$\pm$2.9 & 15.9$\pm$1.4 &  4.5$\pm$2.2 &  47.0$\pm$4.1 \\
      hd76341 &  5.6$\pm$2.4 &  6.0$\pm$4.8 & 20.7$\pm$2.5 &  6.4$\pm$4.9 &  64.2$\pm$8.3 \\
      hd78344 & 36.1$\pm$2.0 & 26.7$\pm$2.3 & 73.6$\pm$2.7 & 30.3$\pm$2.9 & 265.1$\pm$3.0 \\
      hd80077 & 44.2$\pm$1.5 & 35.7$\pm$2.4 & 77.4$\pm$1.3 & 35.6$\pm$1.7 & 289.0$\pm$2.5 \\
      hd91824 &  7.1$\pm$2.5 &  3.3$\pm$4.5 & 14.1$\pm$3.7 &  3.2$\pm$4.5 &  47.5$\pm$6.5 \\
    Tr~16~112 &  5.4$\pm$3.8 &  5.3$\pm$4.3 & 19.7$\pm$3.2 &  9.2$\pm$5.3 &  77.8$\pm$8.6 \\
      hd97253 & 11.7$\pm$4.4 & 10.5$\pm$3.5 & 30.2$\pm$3.5 & 14.8$\pm$6.7 & 131.6$\pm$8.9 \\
     hd100099 &     $\cdots$ &  3.9$\pm$2.9 & 20.6$\pm$3.9 &  3.9$\pm$2.9 &  62.8$\pm$9.3 \\
     hd110432 &  6.1$\pm$2.1 &  6.1$\pm$3.4 & 16.2$\pm$1.1 & 11.6$\pm$2.8 &  69.3$\pm$4.3 \\
     hd142468 & 20.8$\pm$1.6 & 15.8$\pm$1.4 & 55.7$\pm$2.3 & 22.5$\pm$1.6 & 176.8$\pm$2.4 \\
     hd144217 &  4.1$\pm$2.7 &  4.9$\pm$6.3 & 11.7$\pm$3.5 &  4.5$\pm$5.4 &  47.2$\pm$9.3 \\
     hd144470 &  3.9$\pm$2.3 &  3.4$\pm$7.3 & 13.6$\pm$3.3 &  6.6$\pm$7.0 &  59.7$\pm$6.6 \\
     hd145502 &  6.8$\pm$2.3 &  3.5$\pm$5.5 & 14.1$\pm$3.5 &  6.3$\pm$4.3 &  57.5$\pm$7.6 \\
      sig~Sco &  3.5$\pm$1.6 &  0.9$\pm$0.5 & 16.4$\pm$1.5 &  5.5$\pm$1.4 &  57.8$\pm$3.4 \\
     hd147888 & 10.4$\pm$3.2 &     $\cdots$ & 17.8$\pm$3.3 &  7.5$\pm$5.5 &  72.8$\pm$10.9 \\
     hd147889 & 26.8$\pm$3.5 & 16.6$\pm$3.0 & 34.0$\pm$3.7 & 25.0$\pm$3.2 & 178.9$\pm$6.2 \\
     hd147933 & 10.8$\pm$3.5 &  4.2$\pm$3.0 & 15.0$\pm$2.1 &  8.4$\pm$2.3 &  62.8$\pm$6.4 \\
      chi~Oph &  6.6$\pm$2.1 &  2.4$\pm$1.2 & 12.0$\pm$1.8 &     $\cdots$ &  40.4$\pm$4.0 \\
     hd148688 & 12.9$\pm$3.1 & 10.4$\pm$4.0 & 36.9$\pm$1.9 & 10.4$\pm$2.1 & 121.2$\pm$6.3 \\
      zet~Oph &  3.4$\pm$1.7 &  3.0$\pm$2.2 &  8.3$\pm$1.0 &  3.7$\pm$2.0 &  35.0$\pm$2.8 \\
     hd151932 & 11.7$\pm$3.3 &     $\cdots$ & 29.9$\pm$1.8 & 11.9$\pm$2.8 & 114.7$\pm$4.6 \\
     hd152003 & 13.0$\pm$4.7 &  9.1$\pm$3.7 & 28.8$\pm$2.9 & 12.6$\pm$5.2 & 102.9$\pm$5.7 \\
     hd152233 &  9.4$\pm$2.7 &  7.8$\pm$3.5 & 19.3$\pm$1.7 &  9.7$\pm$3.0 &  78.9$\pm$5.6 \\
     hd152235 & 14.7$\pm$2.0 &  9.9$\pm$2.0 & 32.9$\pm$1.2 & 15.9$\pm$3.0 & 132.2$\pm$3.8 \\
     hd152249 &  9.1$\pm$2.9 &  6.6$\pm$2.9 & 19.6$\pm$1.9 &  3.9$\pm$3.0 &  75.6$\pm$5.0 \\
     hd152270 &  7.9$\pm$3.9 & 10.4$\pm$6.7 & 20.6$\pm$2.5 &  8.3$\pm$2.3 &  62.0$\pm$5.8 \\
     hd154090 &  6.1$\pm$2.7 &  6.3$\pm$1.5 & 18.8$\pm$1.8 & 10.7$\pm$2.4 &  90.5$\pm$5.5 \\
     hd154445 & 10.7$\pm$1.9 &  7.7$\pm$1.8 & 22.3$\pm$1.4 &  8.7$\pm$1.5 & 105.1$\pm$4.8 \\
     hd155450 &  8.6$\pm$4.2 &  3.5$\pm$0.9 & 15.8$\pm$2.1 &  8.2$\pm$2.7 &  67.7$\pm$5.8 \\
     hd159176 &  7.1$\pm$1.0 &  7.1$\pm$1.8 & 20.1$\pm$1.4 &  7.7$\pm$1.8 &  63.1$\pm$2.6 \\
     hd161056 & 23.1$\pm$3.9 & 18.0$\pm$4.8 & 32.6$\pm$2.3 & 17.5$\pm$3.4 & 152.4$\pm$6.3 \\
    V2052~Oph & 10.4$\pm$2.3 &  7.7$\pm$4.0 & 25.5$\pm$2.7 & 15.3$\pm$5.2 & 115.0$\pm$9.0 \\
     hd163800 & 18.4$\pm$2.4 & 11.4$\pm$3.2 & 27.3$\pm$1.4 & 12.2$\pm$2.8 & 119.1$\pm$4.9 \\
        9~Sgr &  7.5$\pm$3.9 &  6.1$\pm$3.9 & 15.6$\pm$2.6 &  7.4$\pm$4.7 &  53.3$\pm$6.1 \\
       15~Sgr & 14.8$\pm$2.6 &  6.1$\pm$5.2 & 21.9$\pm$3.8 &  9.8$\pm$7.8 &  86.3$\pm$8.4 \\
     hd167771 &     $\cdots$ &  7.2$\pm$8.1 & 16.9$\pm$3.4 &  6.4$\pm$2.9 &  61.1$\pm$5.4 \\
     hd169454 & 27.8$\pm$2.9 & 20.8$\pm$3.5 & 51.6$\pm$3.1 & 25.3$\pm$4.9 & 190.5$\pm$4.6 \\
     hd170740 & 13.2$\pm$3.2 &  7.3$\pm$4.9 & 24.3$\pm$2.9 & 13.0$\pm$5.0 & 118.8$\pm$10.6 \\
       20~Aql & 11.0$\pm$2.5 &  7.5$\pm$2.2 & 19.4$\pm$1.6 &  6.4$\pm$1.3 &  92.0$\pm$4.3 \\
     hd184915 &  4.7$\pm$1.8 &  5.2$\pm$2.6 & 16.0$\pm$1.9 &  8.0$\pm$2.9 &  69.1$\pm$5.3 \\
     hd185418 & 12.7$\pm$1.6 & 15.3$\pm$5.4 & 34.5$\pm$2.4 & 15.0$\pm$3.2 & 164.9$\pm$6.9 \\
     hd185859 & 28.9$\pm$2.9 & 21.0$\pm$5.4 & 42.6$\pm$2.5 & 25.6$\pm$8.8 & 206.0$\pm$11.1 \\
     hd193793 & 12.5$\pm$2.3 & 12.7$\pm$4.3 & 43.0$\pm$2.4 & 16.4$\pm$3.9 & 137.4$\pm$4.0 \\
     hd203532 &  8.8$\pm$2.2 &  4.6$\pm$3.8 & 12.2$\pm$2.2 &  4.6$\pm$4.8 &  56.8$\pm$10.1 \\
     hd228712 & 19.7$\pm$4.1 & 17.2$\pm$5.7 & 53.6$\pm$5.1 & 24.0$\pm$5.4 & 180.7$\pm$6.7 \\
\hline
\end{tabular}
\end{table*}

Measurements of the equivalent widths with their 1$\sigma$
uncertainties for the five diffuse bands are given in the
Table~\ref{dib_ew}. The uncertainties were calculated for pure photon
noise statistics by technique used in work by \cite{Vo06}. It remains
indefinite how much such factors as continuum placement, peculiarities
in profiles of diffuse bands themselves (wings, asymmetry); stellar
peculiarities~--- binarity, variability (see the Table~\ref{targets})
which provoke blends with DIBs and also telluric lines remnants, in
case of DIB 6445 \AA, can contribute to the errors in \ew s.  

Correct S/N determination can also present some difficulties. DIBs
6113, 6445 and especially 5545 \AA\ have closely located neighbours
and thus too short pieces of free continuum to measure S/N. So, if
those pieces of spectra were inappropriate for measurements, S/N was
to be measured somewhere in other place of the spectral order. 

\subsection{Correlations between the DIBs.}

\begin{figure*}
\centering
  \begin{tabular}{@{}cc@{}}
    \includegraphics[width=.375\textwidth]{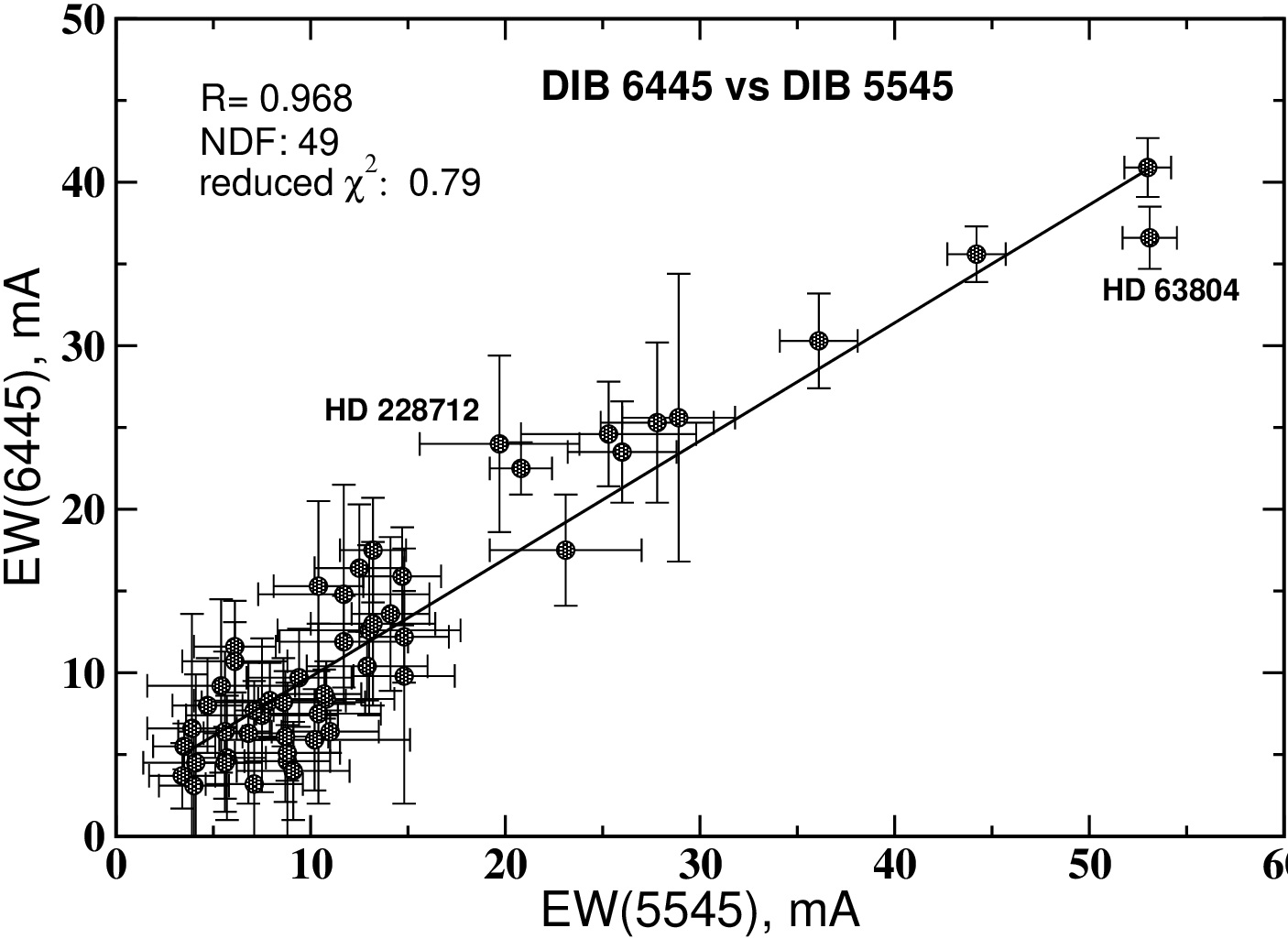} &
    \includegraphics[width=.375\textwidth]{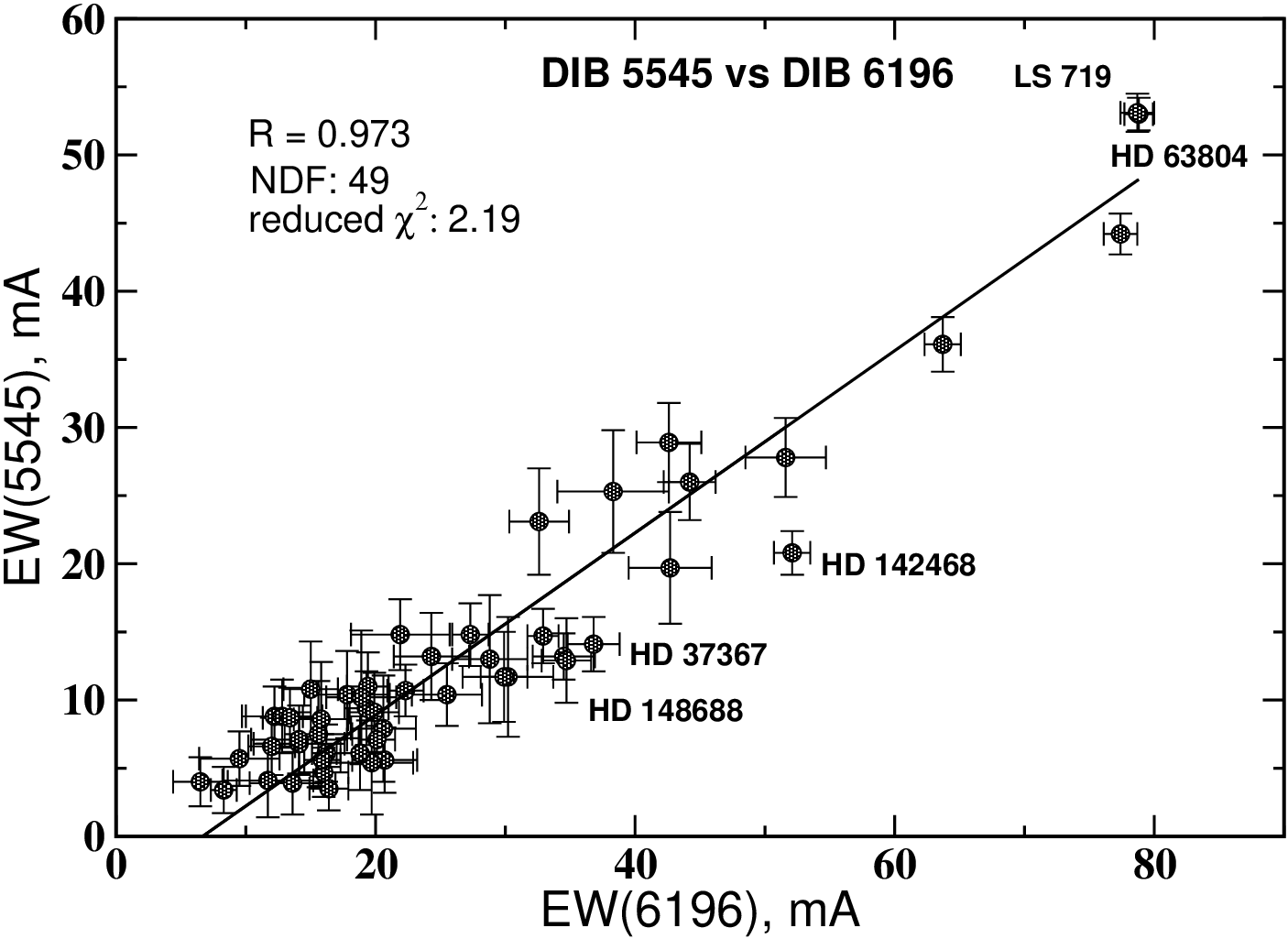} \\
    \includegraphics[width=.375\textwidth]{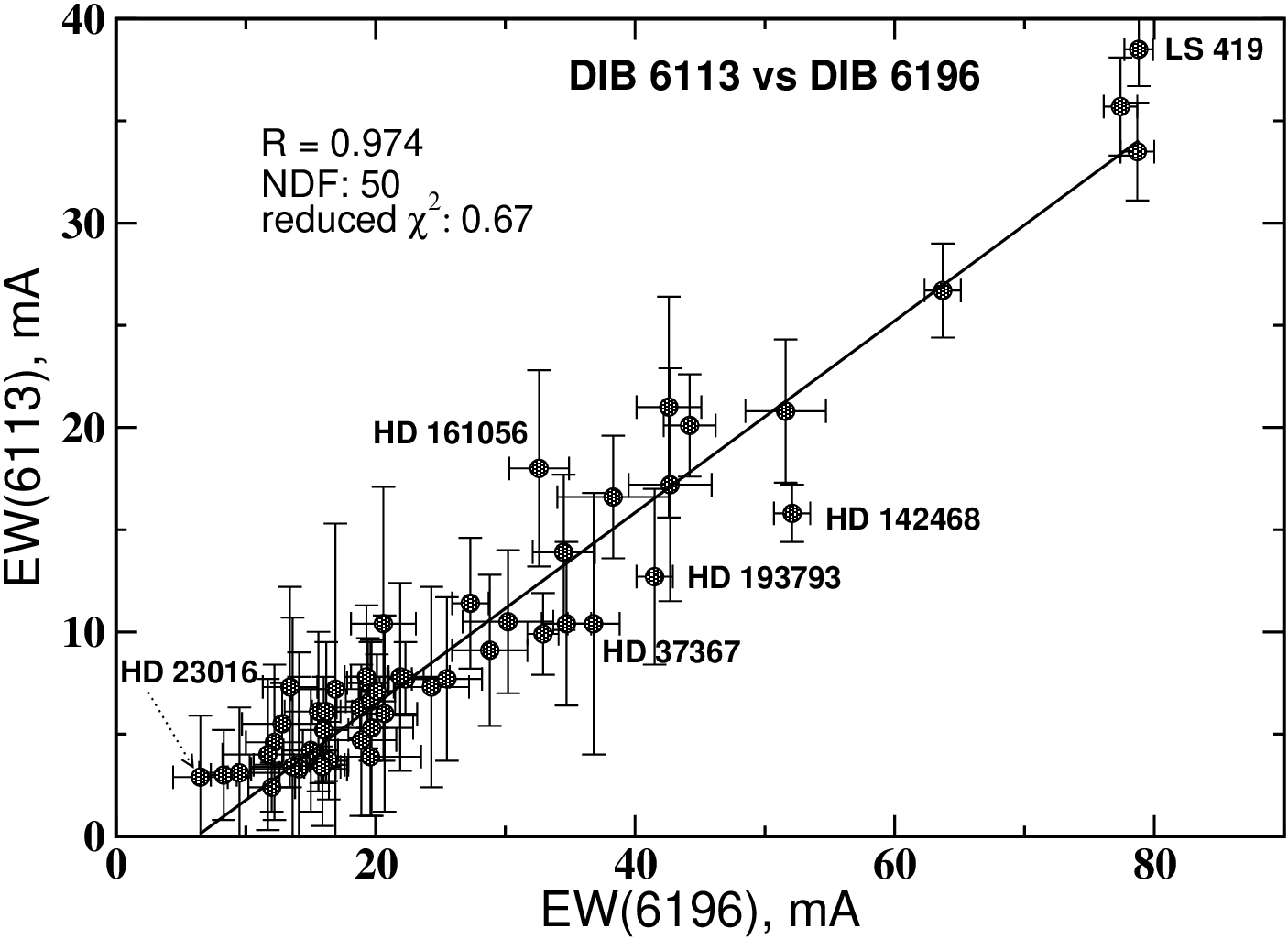} &
    \includegraphics[width=.375\textwidth]{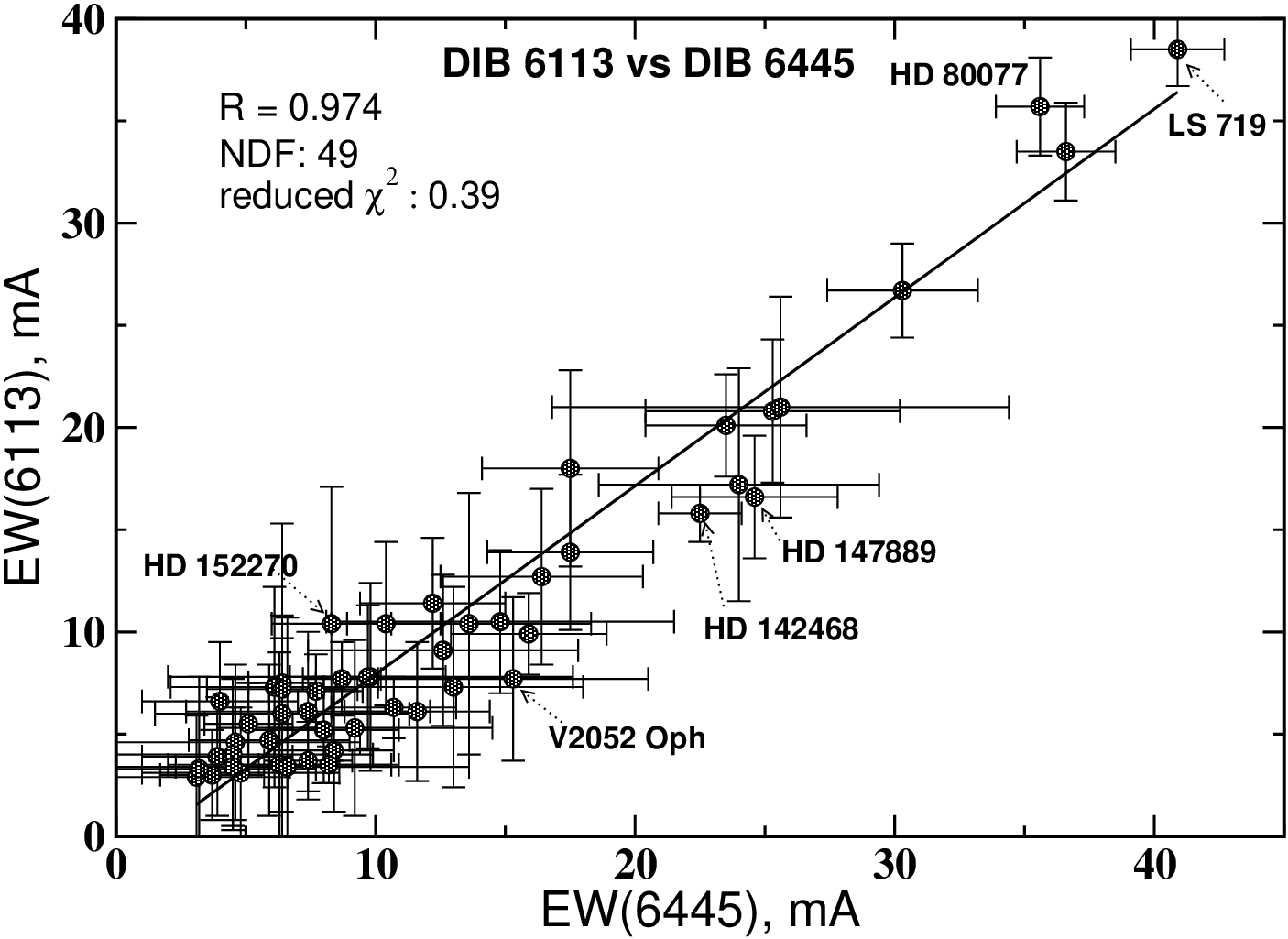} \\
    \includegraphics[width=.375\textwidth]{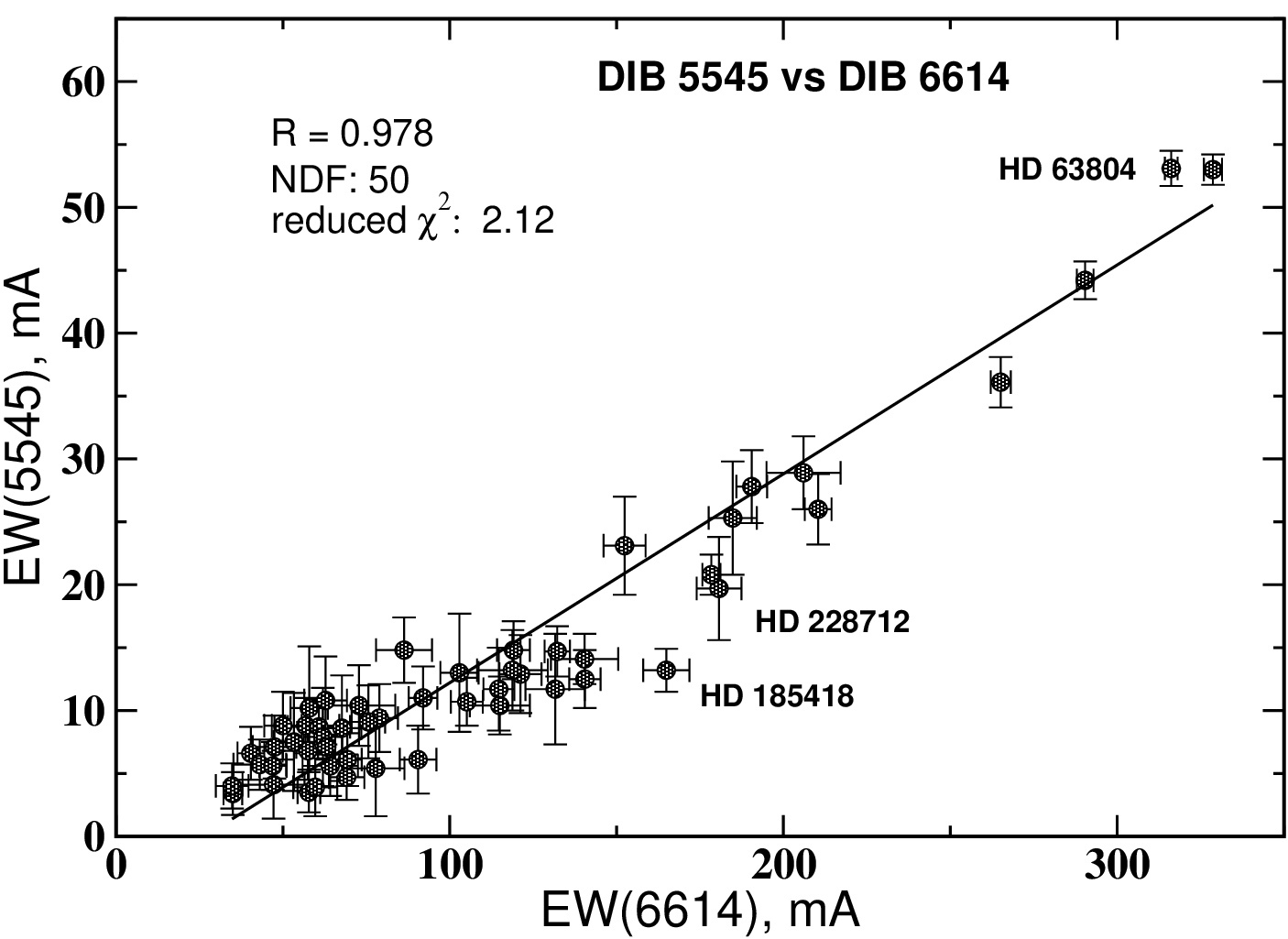} &
    \includegraphics[width=.375\textwidth]{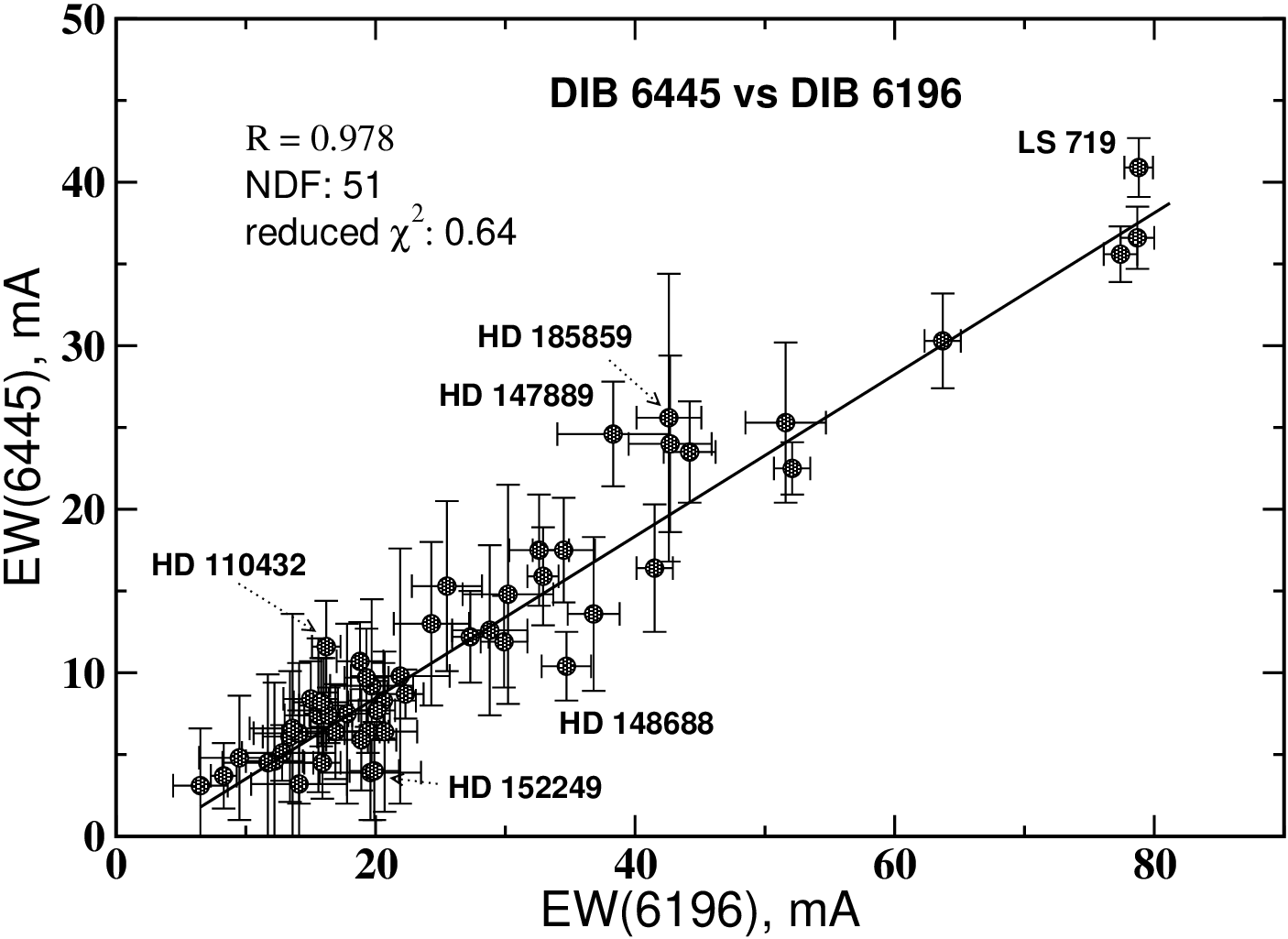} \\
    \includegraphics[width=.375\textwidth]{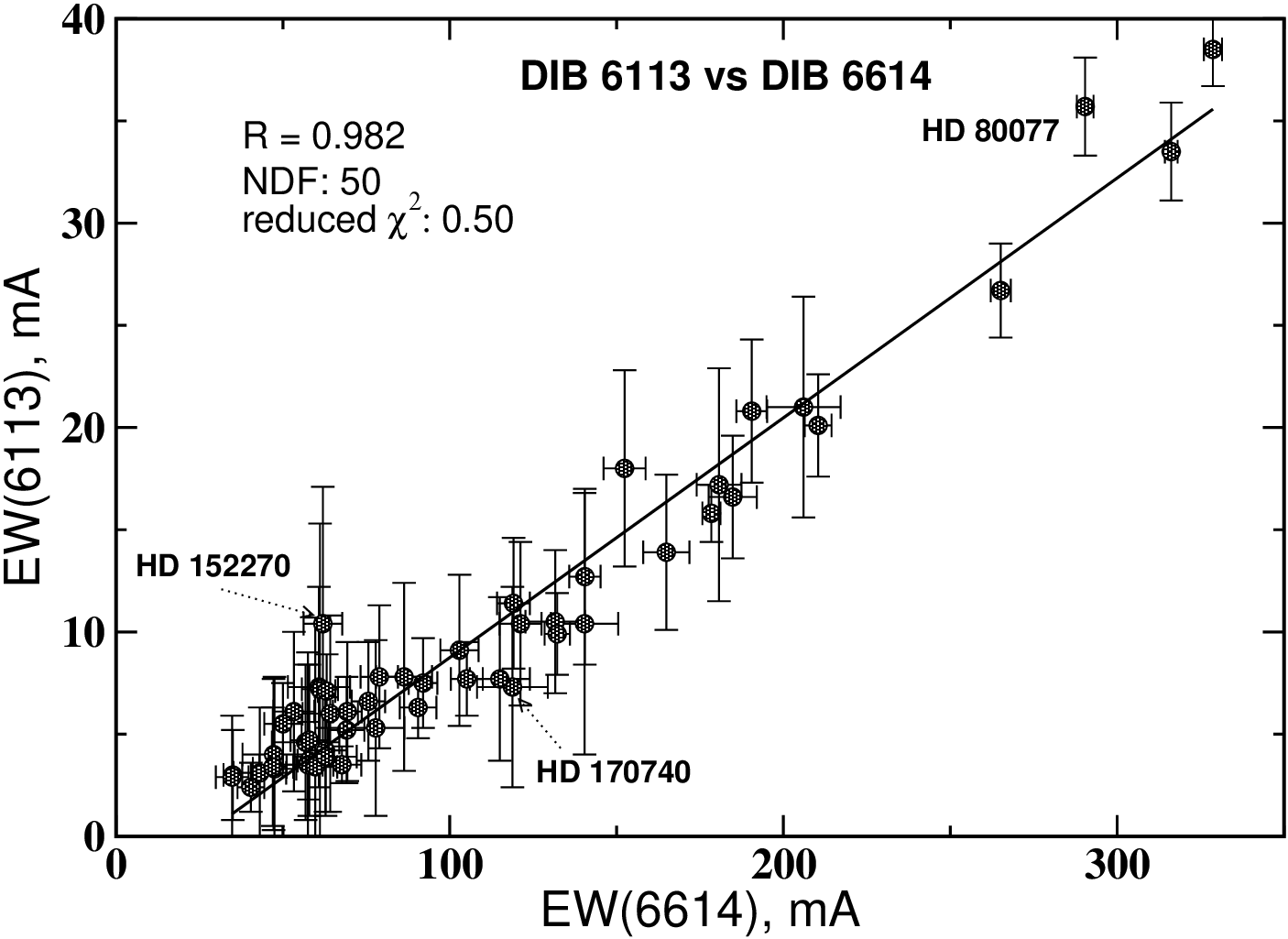} &
    \includegraphics[width=.375\textwidth]{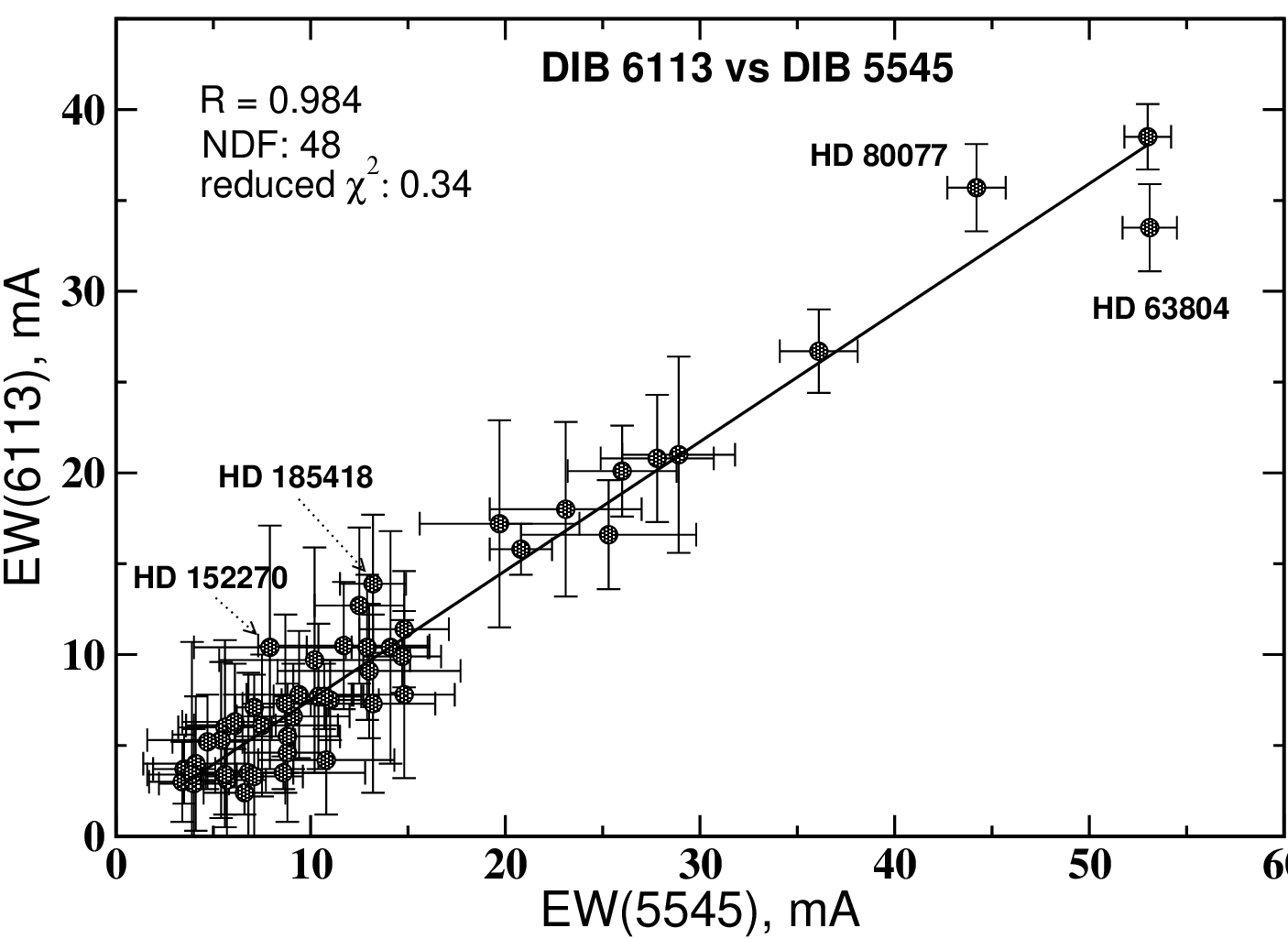} \\
    \includegraphics[width=.375\textwidth]{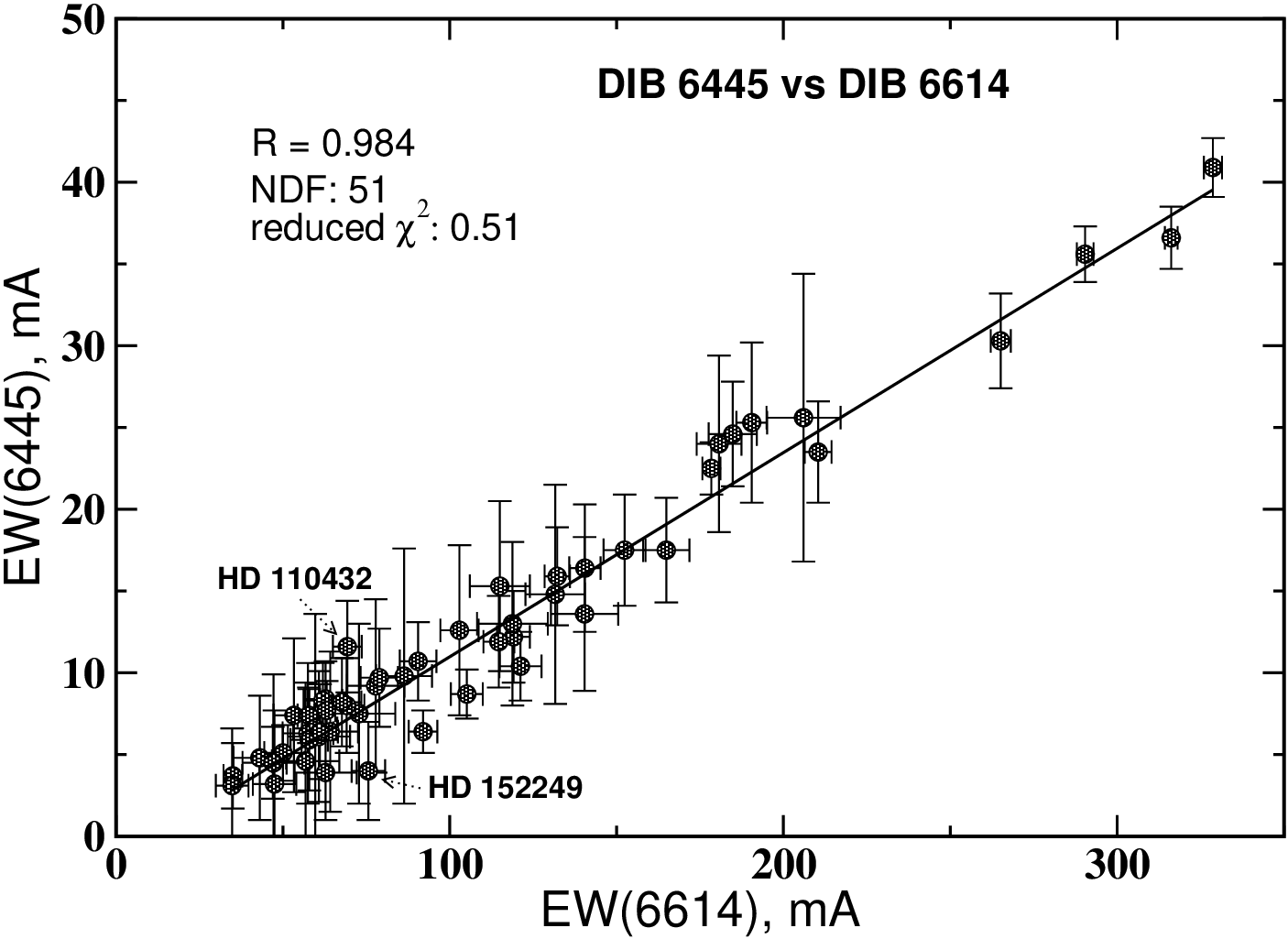} &
    \includegraphics[width=.375\textwidth]{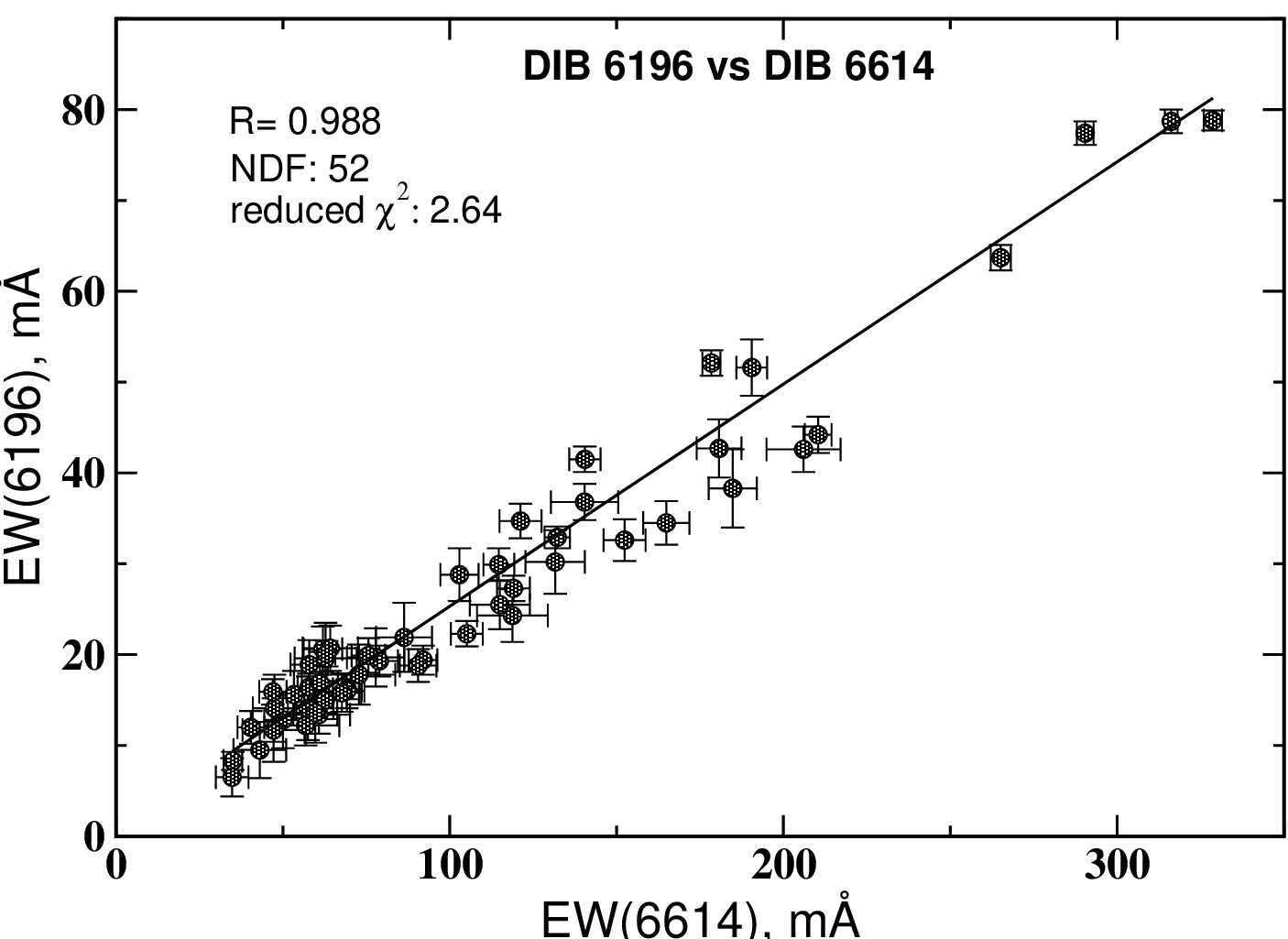} \\
  \end{tabular}
\caption{Weighted fits of the selected DIBs, ordered on increasing 
linear correlation coefficient R. NDF is short for ``Number Degree of
Freedom''. In most cases the deviant points likely are caused by blends of
the DIBs with stellar lines. Low values of $\chi^2 < 1$ may be due to
big uncertainties in the \ew s of two the weakest DIBs: 6113 and 6445
\AA.}
\label{wfits}
\end{figure*}

Interrelations between the DIBs have been sought by weighted least
squares fitting of measured \ew s. All the fits were done by method
described in work by \cite{Or82}~--- case when errors in two variables
were allowed for. Parameters of the fits $a, b$ as well as those for
the inverted fits ($a', b'$) are collected in the Table~\ref{fitspar}.
The direct fits are depicted at the Fig.~\ref{wfits}. In all plots
presented there \ew s of stronger DIB were adopted as arguments (X).
Respectively, to every pair of the DIBs a weighted linear correlation
coefficient, R, has been calculated. The values of R, the number
degree of freedom of the fit (NDF) and reduced $\chi^2$ are shown at
the upper left corners on the plots at the Fig.~\ref{wfits}. 

As it is seen from the figure, the highest value of R = 0.988
was obtained for the pair: DIB 6196 vs DIB 6614~\AA. The correlation
coefficient R, parameters $b = 0.25\pm0.005$ and $b' = 3.99\pm0.09$
determined from the direct and inverse fits between EW(6196) and
EW(6614) (see the Table~\ref{fitspar})~--- are in a good agreement
with those from the earlier works \citep[see][]{MC10,Kr16}. It
concerns especially the former one, where detailed statistical
analysis was done with data set twice larger than in the present work.

Other pairs of the diffuse bands: 6445 vs 6614 \AA, 6113 vs
5545 \AA\ and 6113 vs 6614 \AA\ also show quite a high degree of
correlation R = 0.982~-- 0.984, see the Fig.~\ref{wfits}. Correlations
between other bands remain at nearly equal and still high level R~$>$
0.970. Only one pair~--- DIB 6445 vs DIB 5545 \AA\ has R $<$ 0.970.  

\begin{table}
\caption{Parameters $a,\; b,\; a',\; b'$ obtained from direct and
inverted least squares linear fits of intensities of the selected DIBs.
Direct fits are presented on the plots at the Fig.~\ref{wfits}.}
\label{fitspar}
\begin{tabular}{lrr}
\hline
Pair of DIBs & $a,\quad b$ & $a',\quad b'$ \\
\hline
5545:6196    & $-4.46\pm0.92$ & $7.11\pm1.15$ \\
             & $0.67\pm0.02$ &  $1.45\pm0.05$ \\
5545:6614    & $-4.40\pm0.81$ & $30.79\pm4.11$ \\
             & $0.17\pm0.01$ &  $5.77\pm0.17$ \\
6113:5545    & $0.41\pm0.40$ & $-0.09\pm0.56$ \\
             & $0.71\pm0.01$ &  $1.37\pm0.04$ \\
6113:6196    & $-2.92\pm0.55$ & $6.64\pm0.99$ \\
             & $0.47\pm0.02$ &  $2.06\pm0.07$ \\
6113:6445    & $-1.30\pm0.55$ & $2.00\pm0.53$ \\
             & $0.92\pm0.03$ &  $1.01\pm0.03$ \\
6113:6614    & $-2.99\pm0.45$ & $28.46\pm3.18$ \\
             & $0.12\pm0.00$ &  $8.23\pm0.22$ \\
6196:6614    & $0.85\pm0.86$ & $-0.28\pm3.50$ \\
             & $0.25\pm0.01$ &  $3.99\pm0.09$ \\
6445:5545    & $2.53\pm0.61$ & $-2.44\pm0.91$ \\
             & $0.72\pm0.03$ &  $1.31\pm0.05$ \\
6445:6196    & $-1.43\pm0.56$ & $3.56\pm1.03$ \\
             & $0.49\pm0.02$ &  $1.96\pm0.06$ \\
6445:6614    & $1.51\pm0.47$ & $15.42\pm3.46$ \\
             & $0.13\pm0.00$ &  $7.76\pm0.20$ \\
\hline
\end{tabular}
\end{table}

It is seen from the 2nd and 3rd columns of the Table~\ref{fitspar}
that the ratios of \ew s probably demonstrate a sequence: {\small
EW(6196)/EW(6614)$\,\approx$1:4, EW(5545)/EW(6614)$\,\approx$1:6, and
pairs EW(6445)/EW(6614), EW(6113)/EW(6614)$\,\approx$1:8}. 

In vast majority of cases the most deviant points at the
Fig.~\ref{wfits} could be explained due to blends with stellar lines.
So, in accordance with the synthetic spectra generated with above
mentioned programs, ATLAS9 and SYNTHE, DIB 5545 \AA\ in the spectrum
of B9.5Ia star HD\,63804 (see the plots on the Fig.~\ref{wfits}) looks
overestimated probably due to the blend with the stellar \ion{Fe}{II}
line. And further: DIB 6113~\AA\ in the spectra of HD\,185418 (B0.5V),
HD\,152270 (WC7+O5-8), LS\,719 (B3Ia) and HD\,80077 (B2Ia) probably
has the blend with \ion{Ne}{II} or with \ion{Fe}{II} in the spectrum
of HD\,23016 (B9Vne). DIB 6445 \AA\ can be blended with the stellar
\ion{Fe}{II} line in the spectra of V2052~Oph (B3III), HD\,147889
(B2III/IV) or with \ion{N}{III} lines in the spectra of HD\,110432
(B0.5IVpe), HD\,142468 (B1Ia/Iab), HD\,185859 (B0.5Iae), HD\,228712
(B0.5Ia). As for DIB 6196 \AA, e.\,g in the spectra of HD\,37367
(B2IV-V), HD\,142468 (B1Ia/Iab) or other stars with the deviant points
at the Fig.~\ref{wfits}, it is difficult to say what could increase
the DIB's \ew\ as the stellar \ion{Fe}{III}, near $\lambda$
6194.8~\AA, is located quite far from the DIB to create a blend.
Contrary can be observable for DIB 5545~\AA\ in the spectrum of
HD\,185418, see the plots EW(5545) vs EW(6614) or EW(6113) vs
EW(5545), there the meaning of \ew\ is much lower of that predicted by
fit.  

So, in some cases neither too big nor too small values of \ew s could
be explained. If these effects have not been caused by cosmic ray hits
or by defects in spectra then specific local conditions of ISM might
be responsible. It is quite probable that such deviant points
could be caused by variations in the local physical conditions,
especially in the intensity of radiation field and in the interstellar
density, which can influence DIB equivalent widths and shapes of their
profiles. This effect can be seen, for instance, in spectra of stars
in Upper Scorpius \citep{Vs11} and in spectra of Local Bubble objects
\citep{Ba16, Fh19} or in the spectrum of Her~36 where DIBs have
extended red wings \citep{Da13, Ok13}. 

In many cases weakness of DIBs may be really explained by diminishing
of DIB carriers in interstellar medium because of proximity of the hot
stars, as it is in Orion Nebula. From the other hand DIBs can be more
intense or of normal strengths under nearly the same physical
conditions \citep[see][]{Hb95}. 

\subsection{Spacings between the DIBs.}

The DIBs explored here make a set of wavenumber differences or
spacings, \deltnu, which cover the range 218~-- 2913~\icm, see the 2nd
column of the Table~\ref{spacings}. These spacings pretty well match
to the energies of vibrational transitions in some molecules including
polyatomic ones, such as PAHs, which are quite abundant in ISM
\citep[see][]{Lg84}. Such transitions, often ascribed to vibrations in
PAHs, are observable as IR emission bands, named earlier as
Unidentified Infra-Red (UIR) emission features, see, e.\,g. review by
\cite{Mt90} and references therein. Several IR bands at
$\lambda\lambda$ 16.4, 11.3, 7,7, 6.2 and 3.3 \micron\ are presented
in the 3rd column of the Table~\ref{spacings}. The IR band at
16.4~\micron\ is described in works by \citep{Bl98,Te99,Mo00} about
four others see, e.\,g. \citep{Lg84,Mt90}. In addition there is one
band, at 5.3~\micron, visible in calculated spectra of PAHs
\citep[see][]{Lg89}. 

\begin{table}
\caption{Spacings between the DIBs in comparison with wavenumbers
of some IR bands ascribed to vibrations in some PAHs.}
\label{spacings}
\begin{tabular}{lrrc}
\hline
DIBs pair & \deltnu$^a$, & IR band$^b$, & Ratio$^c$ \\
& \icm\ & \tnu, \icm\ ($\lambda$, \micron) & \\ 
\hline
6196 $\leftrightarrow$ 6113 & 218.5  & $\cdots$    &$\cdots$  \\
6614 $\leftrightarrow$ 6445 & 394.9  & $\cdots$    &$\cdots$  \\
6445 $\leftrightarrow$ 6196 & 624.1  & 609.8 (16.4)&1.02      \\
6445 $\leftrightarrow$ 6113 & 842.5  & 884.8 (11.3)&0.95      \\
6614 $\leftrightarrow$ 6196 & 1018.9 & $\cdots$    &$\cdots$  \\
6614 $\leftrightarrow$ 6113 & 1237.4 & 1298.4 (7.7)&0.95      \\
6113 $\leftrightarrow$ 5545 & 1675.8 & 1612.9 (6.2)&1.04      \\
6196 $\leftrightarrow$ 5545 & 1894.3 & 1886.8 (5.3)$^d$&1.00  \\
6445 $\leftrightarrow$ 5545 & 2518.3 & $\cdots$    &$\cdots$  \\
6614 $\leftrightarrow$ 5545 & 2913.2 & 3029.7 (3.3)&0.96  \\
\hline
Notes. \\
\multicolumn{4}{p{200pt}}{$^a$ --- spacings between DIBs.} \\
\multicolumn{4}{p{200pt}}{$^b$ --- wavenumbers and wavelengths
of vibronic infrared bands ascribed to PAHs.} \\
\multicolumn{4}{p{200pt}}{$^c$ --- the ratio spacing/wavenumber of IR
band (\deltnu/\tnu).} \\
\multicolumn{4}{p{200pt}}{$^d$ --- the band observed in calculated 
spectra for PAHs, see \cite{Lg89}.} 
\end{tabular}
\end{table}

Of all spacings collected in the second column of the
Table~\ref{spacings} six~--- with \deltnu~= 624.1, 842.5, 1237.4,
1675.8, 1894.3 and 2913.2 $cm^{-1}$ --- demonstrate a relation with
the wavenumbers of IR bands. Thus spacings 842.5, 1237.4, 1675.8 and
2913.2 \icm\ are lesser/greater than their counterparts from IR bands
by nearly equal factors (1.04, 1.05), see the 4th column of the
Table~\ref{spacings}. These IR emissions is now regarded as
indicatives of aromatic hydrocarbons due to CH out-of-plane bend
vibration at $\lambda$ 11.3 \micron\ (\tnu\ = 884.8~\icm), C~---~C at
$\lambda\lambda$~6.2 \micron\ (1612.9 \icm), 7.7~\micron\
(1298.4~\icm) and CH at $\lambda$ 3.3~\micron\ (3029.7~\icm) stretch
vibrations, see, for instance, review by \cite{Mt90}. The spacing
\deltnu~= 624.1~cm$^{-1}$ matches (it is greater by factor 1.02) to
the band at $\lambda$~16.4~\micron\ (609.8~\icm) which a PAH produces
due to vibrations \citep[see][]{Mo00}. This IR emission has been known
from laboratory experiments \citep{Mo96} as 16.2~\micron\ mode in PAH
and was first mentioned in \cite{Bl98} as emission observable at
$\lambda$ 16.5~\micron\ in spectra of NGC\,7023 and $\rho$~Oph
molecular cloud. The spacing \deltnu~= 1894.3~\icm\ nearly exactly
corresponds to IR band at 5.3~\micron\ (1886.8 \icm) which was
observed in calculated spectra of PAHs and recommended for search in
astronomical objects by the authors \citep[see][]{Lg89}.

\begin{figure*}
\centering
\includegraphics[width=\textwidth,height=280pt]{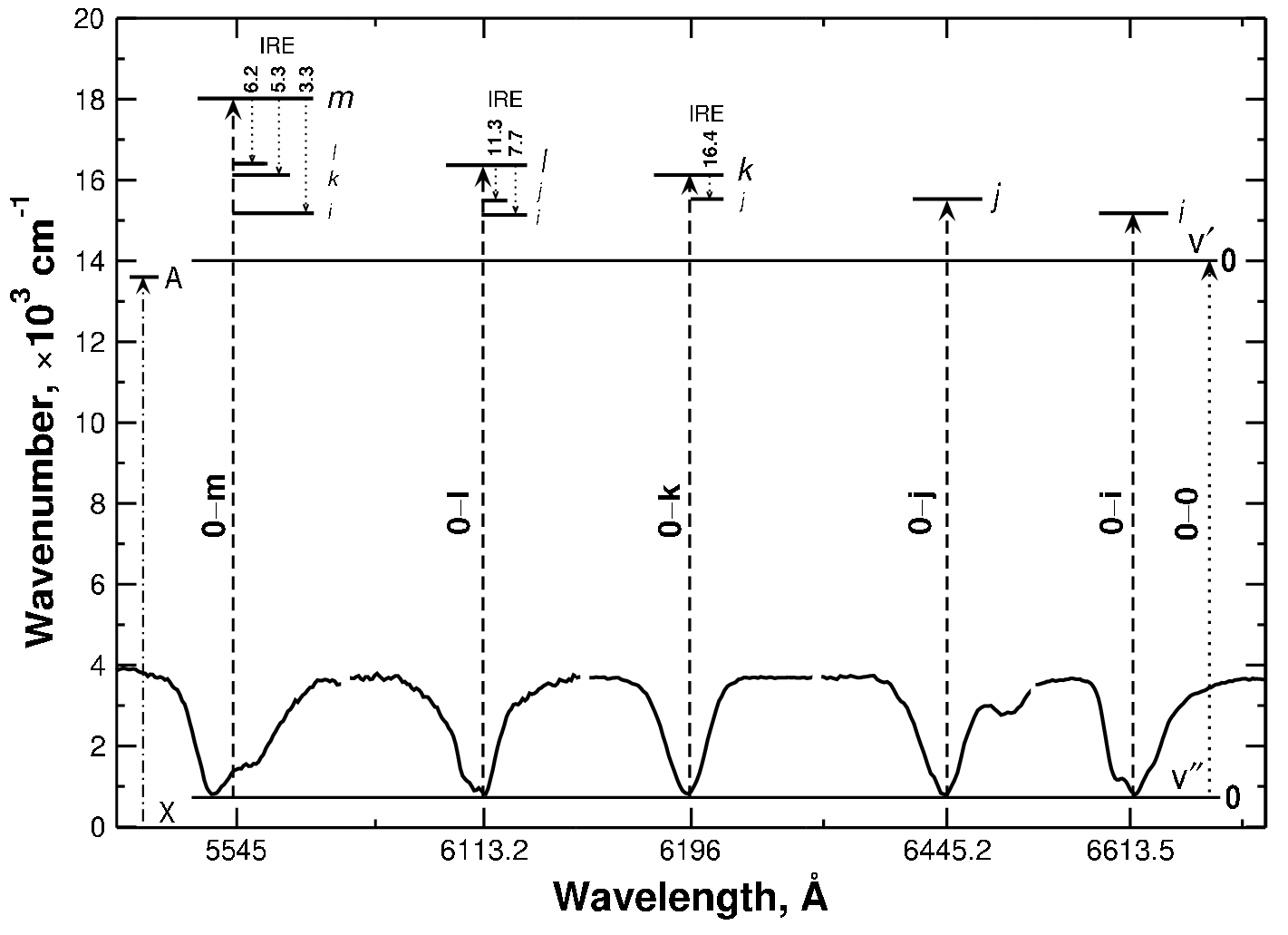}
\caption{Quasi energy-level diagram. Energy of the upper level above
the ground state in 10$^3$\icm. Electronic transition $X \rightarrow
A$ causes vibrational-rotational transitions from $v'' = 0$ to
arbitrary upper levels, that produces the system of DIBs shown below.
Rotational bands are invisible at this scale and probably reveal
themselves as substructures in the DIB profiles. Likely transitions
from upper to lower vibrational states with ensuing IR emissions (IRE)
due to vibrations in some PAHs are marked by numbers in \micron, see
the Table~\ref{spacings}. Radiationless transitions from upper to
lower $v'$ levels (vibrational relaxation) are also possible.}
\label{qenerg}
\end{figure*}

So, four pairs: \deltnu~vs \tnu\ from the Table~\ref{spacings} show
the same $\approx$~5\% and 4\% differences; one pair differs by 2\%
and one by $\approx$ 0.3\%. If to guess PAHs are species responsible
for DIBs, what could induce such a divergence? At the present moment
it is known about variations in the relative strengths as well as
shifts in wavelengths of the IR bands. They may arise from
changes in PAHs mixture, including changes in fraction of positively
or negatively charged PAHs \citep{Lg89,Mt90}. Modes of vibrations
depend on internal structure of molecule. For instance, in laboratory
spectra the band at 16.4~\micron\ can be produced in the wavelength
range 16.21~-- 16.52~\micron\ (\tnu~= 623~-- 611~\icm), thus varying
in its position $\approx$ by 2\% \citep{Mo00}. The band at
11.3~\micron, caused by out-of-plane C~---~H bending occurs at
$\lambda\lambda$ 11.6~-- 12.5~\micron\ (\tnu\ = 862~-- 800 \icm) if
there two C~---~H bonds on the same aromatic ring, and at $\lambda\lambda$
12.4~-- 13.3~\micron\ (807~-- 752 \icm) for three bonds (see
\cite{Mt90} and references therein). \cite{Te99} demonstrated IR bands
in the Red Rectangle with $\lambda\lambda$ 11.0, 11.3, 11.9,
12.8~\micron. According to the Table~\ref{spacings} the band at
11.9~\micron\ (840.3~\icm) quite well corresponds to the spacing
\deltnu\ = 842.5~\icm. 

Wavelengths of IR bands also depend on type of their carriers:
whether they are in aromatic or in aliphatic forms. In PAHs C~---~H
stretching gives a band at $\lambda$ 3.3 \micron. Aliphatic
hydrocarbons have a band of weaker intensity at $\lambda$ 3.4~\micron\
(2941.2~\icm) which even better matches to the spacing \deltnu\ =
2913.2 \icm, from the Table~\ref{spacings}.  According to the recent
studies \citep[see e.\,g.][]{Li12} the fraction of aliphatic
constituent in IR bands carriers is $<$ 15\%~--- thus they are mostly
aromatic. As to the possible carriers of diffuse bands, it is
improbably that this group of five DIBs was due to both aliphatic and
aromatic hydrocarbons.

In any case a found proximity of spacings between the DIBs to
wavenumbers of several IR bands may point to common origin of their
carriers: be they PAHs or any other complex molecules. 

If to suppose that the five diffuse bands analysed in the current
study present a fragment of spectrum of a complex molecule~--- e.\,g.
an electronic transition with rotation-vibrational series~--- then DIB
profiles could be considered as packed, unresolved rotational
structure and spacings between the DIBs as differences in energies of
vibrational states. Such a series perhaps might look as it is
depicted at the Fig.~\ref{qenerg}. Electronic transition from the
ground level to an upper state $X \rightarrow A$ (e.\,g. after
absorption an energetic photon) can give rise to a series of
vibrational transitions from the ground vibrational level $v'' = 0$ to
arbitrary upper levels $v' =$ i, j, k, l, m etc. with a set of
rotational transitions. Thus, the DIBs at corresponding wavelengths may
appear. Later on due to different intramolecular processes the system
can go to lower energy states in different ways. One of them
vibrational relaxation~--- energy transfer from upper vibrational
levels to different modes of vibrations (radiationless transition
without changing electronic state), another~--- vibrational
de-excitation through IR emissions.  Several likely transitions
between upper ($v'$) vibrational levels which could lead to IR
emissions are shown at the Fig.~\ref{qenerg}, see also the
Table~\ref{spacings}. Rotational P, Q, R branch structure may reveal
itself as substructure in DIB profiles, see the Fig.~\ref{profil}.
Especially it is evident in DIB 6614 \AA\ \citep[see][]{Sa95,Ca04}.
Sometimes a similar substructure (as well as the profile broadening)
is developed in DIBs when their carriers are located in regions with
higher temperature, determined e.\,g. on excitation temperature of
C$_2$, as it was shown for DIB 6196~\AA\ in the study by \cite{Kz09}.
On the Fig.~\ref{profil} the substructure is evident in the profile of
DIB 6196~\AA\ in the spectrum of HD\,147889. Conspicuous substructure
is also observable in the profile of the band at 5545~\AA.

If the five DIBs appear as a result of above mentioned factors in such
a case the intensity ratios of the DIBs, noted in subsection 3.3~---
EW(6196)/EW(6614)$\,\approx$1:4, EW(5545)/EW(6614)$\,\approx$1:6,
EW(6445)/EW(6614) and EW(6113)/EW(6614)$\,\approx$1:8, see the
Table~\ref{fitspar}, might have been ratios of the Franck-Condon
factors which determine how intensity of an electronic transition is
distributed between the different vibrational bands.

On the whole there is not any evident regularity in the found
spacings. Moreover in complex molecules vibronic progressions do not
necessarily follow simple equation for energies in function of
vibrational quantum number `$v$' for the given fundamental frequency
`$w_e$' and inharmonicity `$x_e$' constants, which depend on
electronic state of the molecule. So, finding a regularity in such a
case may present a difficult task. 

If to look into the whole DIBs' spectrum there are many pairs of DIBs
(not so strong) with values of spacings close to the presented in the
Table~\ref{spacings}.  Those with the best coincidence $\sol$0.1~\icm\
are next: 6600 \lrar\ 5866 (1894.3 \icm); 6486 \lrar\ 5850 (1675.8
\icm); 6768 \lrar\ 6245 (1247.4~\icm); 6325 \lrar\ 6005 (842.5 \icm);
6467 \lrar\ 6216 (624.1 \icm). Many others do not match so well, but
here, of course, the precision of DIB wavelengths should be taken into
account.

\section*{Conclusions}

High linear correlation coefficients R = 0.968~-- 0.988 could indicate
close interrelation of the five considered bands. Correlation and
intensities ratio found for two diffuse bands at $\lambda$ 6196 and
6614~\AA\ are in a good accordance with the result of \cite{MC10}, see
the Fig.~\ref{wfits} and the Table~\ref{fitspar}. Two DIBs $\lambda$
6445 and 6113~\AA\ slightly better correlate (R $>$ 0.98) with DIB
6614~\AA\ than with DIB 6196~\AA; DIB at $\lambda$ 5545~\AA\
correlates worse and almost equally with both~--- R $<$ 0.98. 

Low $\chi^2$ could point to errors underestimation and also result 
from big uncertainties in EW measurements of the weakest DIBs in the
set: 6113 and 6445 \AA.

Do these DIBs have a common origin despite that fact the forms of
their profiles are not alike? There is some similarity in the profiles
of DIBs at $\lambda\lambda$ 6113 and 6196 \AA; and three other bands
have such a common feature as redward extension. 

The spacings presented in the Table~\ref{spacings} do indicate that
DIB carriers might rather be related with PAHs or molecular compounds
which contain groups of molecules such as CH, C$_2$. Five spacings
have values close to the wavenumbers of IR emission bands which were
suggested to arise due to vibrations in PAHs: at $\lambda\lambda$
16.4, 11.3, 7,7, 6.2 and 3.3 \micron; in addition there is one spacing
(\deltnu\ = 1894.3 \icm) that corresponds to the IR band at
5.3~\micron\ (1886.8~\icm), visible in calculated spectra of PAHs
\citep[see][]{Lg89}. 

There is not any evident regularity in the spacings amid the
considered DIBs. But a proximity of the spacings to the wavenumbers of
the IR bands may be a sign of a close relation of the DIBs' carriers
to PAHs or even that the PAHs are these carriers themselves. It seems
that DIBs may arise as a result of excitation of PAHs. IR bands could
appear due to intramolecular processes such as de-excitation of
vibrational levels in PAHs, see the Fig.~\ref{qenerg}. Perhaps, to
shed light on this issue it makes sense to search and explore the
objects in whose spectra both IR bands and DIBs are observable. 

Found ratios of equivalent widths have values:
EW(6196)/EW(6614)$\,\approx$1:4, EW(5545)/EW (6614)$\,\approx$1:6,
EW(6445)/EW(6614) and EW(6113)/EW(6614)$\,\approx$1:8. Whether it is a
sign of a regularity, e.\,g. ratios of their Franck-Condon factors,
which influence energy distribution between vibrational bands in a
complex molecule or simply a coincidence?

This short investigation if not supports the PAH hypothesis of
DIB carriers directly, points to possibility of close relation between
DIB carriers and PAHs \citep[see][]{vZ85, Lg85}. Perhaps a new profound
study of DIBs phenomenon based on PAHs hypothesis may help to answer
the above mentioned questions and to reveal the origin of these mystic
spectral features.

\section*{Acknowledgements}

Based on observations made with ESO Telescopes under programmes ID:
67.D-0439(A), 194.C-0833(A,B,D,F), 382.D-0237, 266.D-5655(A),
67.C-0281(A), 097.C-0979(A), 60.A-9036(A), 076.B-0055(A),
0102.C-0040(B), 080.D-0855(A), 083.C-0503(A), 078.C-0403(A),
187.D-0917(A). The author is very grateful to the referee for his
relevant and useful comments and suggestions.

\bsp
\label{lastpage}
\end {document}